\documentclass[12pt]{JHEP3}
\usepackage{amsmath}
\usepackage{amssymb}
\usepackage{young}
\usepackage[vcentermath]{youngtab}

\newcommand{\be}{ \begin{equation}}
\newcommand{\ee}{\end{equation}}
\newcommand{\bea}[1]{\begin{eqnarray}\label{#1} }
\newcommand{\eea}{\end{eqnarray}}

\def\ZZZ{{\hskip-3pt\hbox{ Z\kern-1.6mm Z}}}
\def\zzz{{\hskip-3pt\hbox{ z\kern-1mm z}}}

\def\Z{{\cal Z}}

\def\one{{\hbox{ 1\kern-.8mm l}}}
\def\zero{{\hbox{ 0\kern-1.5mm 0}}}

\title{String Theory as a Higher Spin Theory}

\author{
Matthias R.\ Gaberdiel$^{a}$ and Rajesh Gopakumar$^{b}$ \\ 
$^a$Institut f\"ur Theoretische Physik, ETH Zurich, \\
$\;$CH-8093 Z\"urich, Switzerland \\
$\;$\email{gaberdiel@itp.phys.ethz.ch}\\ \\ 
$^b$International Centre for Theoretical Sciences-TIFR, \\
$\;$Survey No. 151, Shivakote, Hesaraghatta Hobli, \\
$\;$Bengaluru North, India 560 089\\
$\;$\email{rajesh.gopakumar@icts.res.in}}

\abstract{The symmetries of string theory on ${\rm AdS}_3 \times {\rm S}^3 \times \mathbb{T}^4$ at the 
dual of the symmetric product orbifold point are described by 
a so-called Higher Spin Square (HSS). We show that the massive string spectrum in this background organises itself in terms of 
representations of this HSS, just as the matter in a conventional higher spin theory does so in terms of representations of the 
higher spin algebra. In particular, the entire untwisted sector of the orbifold can be viewed as the Fock 
space built out  of the multiparticle states of a {\it single} representation of the HSS, the so-called `minimal' representation.  
The states in the twisted sector can be described in terms of tensor products of a novel family of representations that are somewhat larger than the minimal one.}

\begin{document}

\section{Introduction}

The consequences of symmetries are much easier to unravel about an unbroken phase rather than, say, a higgsed phase. 
In string theory, while a large underlying symmetry is expected \cite{Gross:1988ue,Witten:1988zd,Moore:1993qe,Sagnotti:2011qp},
we have not been able to pin this down, let alone understand its 
consequences. This is because the backgrounds which we best understand, like flat space, have very few unbroken symmetries. Equivalently, the massless sector of the theory, which reflects the unbroken gauge invariances, has very few states --- those 
of gravity and Yang-Mills (together with their scalar and fermionic partners). 

Through gauge-string dualities we can access other backgrounds which show enhanced unbroken symmetries. The case of 
${\rm AdS}_3 \times {\rm S}^3 \times \mathbb{T}^4$ is one with possibly the largest known unbroken symmetry. The dual 
CFTs are believed to be deformations of a free CFT --- the symmetric product orbifold $(\mathbb{T}^4)^N/S_N$ in the large $N$ limit. 
At the free point they have (as a function of spin) an exponentially large number of single particle conserved currents. These 
correspond to higher spin gauge fields \cite{Gaberdiel:2014cha} in the bulk theory which are `massless' or unhiggsed in this tensionless 
limit \cite{Gaberdiel:2015uca}. At this point in moduli space, the bulk theory contains a Vasiliev higher spin theory \cite{Vasiliev:2003ev} 
as a consistent subsector \cite{Gaberdiel:2014cha}, realising concretely the expectations of 
\cite{Sundborg:2000wp,Witten,Mikhailov:2002bp}.
The underlying higher spin -- CFT duality \cite{Gaberdiel:2013vva} is the supersymmetric generalisation of the original bosonic duality of 
\cite{Gaberdiel:2010pz}, see \cite{Gaberdiel:2012uj} for a review.

In \cite{Gaberdiel:2015mra}, we showed that these single particle currents, which generate the unbroken symmetry 
algebra, have a rather novel underlying structure -- which we dubbed as the Higher Spin Square (HSS). The coinage reflects the fact that the 
commutators amongst arbitrary generators is completely determined in terms of the structure constants of {\it two} independent 
higher spin (HS) algebras called the vertical and horizontal algebras. This is so, even though the HSS is exponentially larger 
than the individual HS algebras (which have a constant number of generators at any given spin $s$). We expect the HSS 
to play a major role in constraining the spectrum and interactions of the theory.

To this end, we investigate in this paper the spectrum of string theory (as captured by the orbifold CFT) in terms of representations 
of both HS and HSS algebras. Our results are as follows. Firstly, we find that the entire untwisted sector of the 
orbifold\footnote{We should stress that, while we 
often refer to the symmetric product orbifold of $\mathbb{T}^4$, in practice, thus far, we have restricted 
to the states which carry zero momentum or winding on the torus. Thus we are effectively considering the symmetric 
product orbifold of $\mathbb{R}^4$.} corresponds 
to the Fock space of multiparticle excitations of a very particular representation of the HSS (together with the 
chiral boundary excitations of the massless 
gauge fields), see eq.~(\ref{2.14a}). This representation is in a sense  the smallest nontrivial representation of the
HSS --- it has the smallest number of states at any given level, or equivalently the largest number of null states.
Furthermore, it is directly built from the so-called minimal representation of the underlying (vertical) HS algebra, which
in turn corresponds to the massive scalar matter field in the higher spin bulk theory. In fact, in the minimal model holography of coset CFTs, the
entire perturbative sector of the higher spin theory is captured, in the dual CFT, by the various tensor powers of the minimal 
representation \cite{Gaberdiel:2011zw}. Thus  the untwisted sector of the CFT dual of  string theory has exactly the same structure with respect to the HSS, as the
perturbative sector of the coset CFTs in the usual minimal model holography. In particular, one would therefore expect that it
is precisely dual to a perturbative higher spin theory (with a suitable matter field) formulated {\it a la} Vasiliev, but now based on the HSS algebra. We emphasise that the HSS is a stringy algebra and its minimal representation is exponentially larger than the corresponding minimal representation of the Vasiliev HS algebra. In fact, the former decomposes into an infinite set of representations of the latter.


Next, we examine the twisted sector of the orbifold. We begin by considering the twisted sector with a single cycle twist of length $m$. This 
is because a twisted sector labelled by a general conjugacy class of $S_N$ with several disjoint cycles of possibly different length can be 
viewed (at large $N$) as multiparticles of the single twist ones. This twisted sector is, in turn, built from separate twisted sectors of the
bosons and fermions with twist $\nu=\frac{r}{m}$, where $r=0, \ldots, m-1$.  We focus largely on a single complex boson, for simplicity, with 
only an abbreviated discussion of the case with more supersymmetry. It turns out that we can view each 
$\nu$-twisted sector as a representation of the underlying vertical HS algebra, thus leading to a novel $1$-parameter family 
(labelled by $\nu$) of representations of the HS algebra. 
With respect to the original HS algebra these 
representations are quite significantly larger than the corresponding minimal representation. This is a reflection of the fact that, as coset representations, they are labelled by large Dynkin labels, as opposed to the minimal representation which is in the fundamental.   
However,  regarded as 
a HSS representation, they are fairly similar in their growth behaviour to the minimal HSS representation. 
The fact that, from the HSS viewpoint, 
the degrees of freedom that arise in the twisted sector have a similar
structure to the degrees of freedom of the untwisted sector is also natural from the string theory 
viewpoint since both describe perturbative degrees of freedom.\footnote{On the other hand, from the 
original HS perspective, the twisted sector representations correspond to 
`non-perturbative'  states \cite{Castro:2011iw, Gaberdiel:2012ku, Perlmutter:2012ds}, and are somewhat mysterious from the 
bulk Vasiliev point of view.} We should mention though that the description of the twisted sector as a representation of the HSS is 
somewhat complicated by the fact
that the generators of the HSS mix the different $\nu$-twisted components in a rather complicated manner;\footnote{A more
positive way of saying this is that the HSS is a Hopf algebra with a non-trivial coproduct.}  we illustrate this effect for the 
simplest example of the $2$-cycle twisted sector.

%

The final upshot is that if we view string theory (on this background) through the lens of higher spin theory, 
the symmetry algebra of the HSS plays the role of the `higher spin' symmetry algebra, 
and  the entire matter content of the theory neatly organises itself in terms of two sets of representations --- 
the minimal and the twisted.  
This seems to suggest that it is very useful to view string theory as a `maxed' out version of a higher spin 
theory with an enormous symmetry algebra and coupling to specific matter field representations. Presumably the 
interactions will also be determined by the symmetry as is largely the case for the Vasiliev theory. Perhaps 
this will allow us to arrive at a unique string field theory (expanded about this background). 
\medskip

The plan of the  paper is as follows. In the next section we briefly review some of the salient features of the HSS,
and give some fairly explicit descriptions of it. In Sec.~3, we describe the untwisted sector and how it organises itself in terms of 
multiparticles of the minimal representation. Sec.~4 studies the twisted sector states as representations of the higher spin algebra. 
We focus mostly on the simpler case of a single complex boson but also make some preliminary remarks on the SUSY cases. 
Sec.~5 goes on to considering the twisted sector states from the point of view of the HSS. The $2$-cycle case is studied in detail,
and some remarks are made about the BPS spectrum (which mainly arises from the twisted sector). 
Sec.~6 ends the paper with some general remarks while the appendices contain some of the 
nitty-gritty details of calculations referred to in the main text.
 
\section{The Higher Spin Square}

Let us begin by reviewing the structure of the unbroken stringy symmetry algebra of the 
${\rm AdS}_3 \times {\rm S}^3 \times \mathbb{T}^4$ background at the point in moduli space described by the symmetric product 
orbifold $(\mathbb{T}^4)^N/S_N$. As discussed in \cite{Gaberdiel:2015mra}, the single particle generators of this algebra are in 
one-to-one correspondence with the chiral algebra of a single supersymmetric $\mathbb{T}^4$ theory, i.e., of four free bosons 
and four free fermions. 
This is basically so since the symmetric product orbifolding corresponds to taking the multiparticles of the individual copies.  
The generating function for the single particle generators is thus
\be
 \sum_{r>0,\, l} \tilde{d}(r,l)\, q^r y^l =
\prod_{n=1}^{\infty} \frac{(1 + y q^{n-\frac{1}{2}})^2 (1 + y^{-1} q^{n-\frac{1}{2}})^2}{(1-q^n)^4}  -1 \ .
\ee
Note that there is an exponentially growing number of states $(\propto \exp{a\sqrt{s}})$ as a function of the spin $s$. This is to be 
contrasted with the constant (independent of $s$) number of gauge fields in a Vasiliev theory. In fact, amongst all these single 
particle generators, only the ones which are built as bilinear combinations of the fermions and bosons appear in the
supersymmetric Vasiliev theory; these generators form a closed subsector of the full algebra, 
as might be expected of a tensionless point of ${\rm AdS}$ string theory \cite{Gaberdiel:2014cha}. 

The idea in \cite{Gaberdiel:2015mra} was to view the complete set of  stringy symmetry generators through the lens of 
this conventional higher spin symmetry. As a first step, the other generators were organised in terms of 
representations of the higher spin subalgebra, and it was found that 
a rather simple set of of representations contribute (see eq.~(2.13) of \cite{Gaberdiel:2015mra}). The next step was 
to try and characterise the stringy algebra entirely in terms of higher spin algebras. This also turned out to be possible, 
essentially because of the presence of an additional, less obvious, higher spin symmetry. This is best illustrated
for the case of the chiral algebra of a single complex boson (rather than the full $\mathbb{T}^4$ theory), and
we shall also, in this paper, often restrict to this case for simplicity. In this case, the generating function for the 
single particle generators equals
\be\label{bosHSS}
\prod_{n=1}^{\infty} \frac{1}{(1- y q^n) (1 - y^{-1} q^n)}  -1 \ .
\ee

For the case of the complex boson, the higher spin algebra (built from the neutral bilinears of these bosons) 
gives rise to the (bosonic) ${\rm hs}[1]$ algebra, whose asymptotic extension equals ${\cal W}_{\infty}[1]$. 
The other single particle generators are then generated by $n$ bosons and $m$ of their complex conjugates, as well as 
an arbitrary number of derivatives on each of these factors. They transform in the representation of ${\rm hs}[1]$ 
labelled by  $(\Lambda_+;\Lambda_-) = ([0^{n-1},1,0,\ldots,0,1,0^{m-1}]; 0)$. These additional representations,
labelled by the pairs $(n,m)$, need to be added to the higher spin algebra ${\rm hs}[1]$ to generate the bosonic analogue 
of the stringy symmetry algebra. We can visualise this by writing a different column for each representation $(n,m)$, with 
the higher spin algebra ${\rm hs}[1]$ corresponding to the first column with $(n,m)=(1,1)$. 
Each such column is generated from the top state $(\partial\phi)^n(\partial \bar{\phi})^m$ upon the action of the ${\rm hs}[1]$ 
generators.

What is interesting is that we can fermionise each real boson ($\phi=\phi+i\phi_2$) into a pair of complex fermions. The 
monomials of the form $(\partial\phi_{1,2})^n$ (together with lower order correction terms) can be written in terms of neutral 
bilinears of the corresponding complex fermion, i.e., as bilinear ${\rm U}(1)\times {\rm U}(1)$ singlets. These bilinears 
in turn generate another higher spin algebra, whose wedge algebra consists of two copies of ${\rm hs}[0]$ (with 
asymptotic extension ${\cal W}_{1+\infty}[0]$). This `horizontal' higher spin algebra gives another way to organise the stringy symmetry
algebra in terms of another set of representations. Together with the earlier `vertical' algebra, these two much smaller 
HS algebras determine recursively all commutators amongst the elements of the stringy symmetry algebra --- which was thus dubbed as a 
`Higher Spin Square' (HSS). While the presence of the two HS algebras is intriguing and is likely to play a significant role in our 
eventual understanding of the structure and consequences of the HSS, it will not enter too much in the present 
paper. We will, in the following, mainly restrict to the more obvious vertical HS algebra, and not utilise the presence of the 
horizontal algebra. 

While this construction should, in principle, determine all commutators of the HSS, the description is somewhat implicit. 
It is therefore of some interest to find a more explicit description of the HSS algebra. In the rest of this section
we shall provide such a description which is somewhat parallel to the familiar oscillator construction for the HS algebras. 
This may therefore play a similar role in uniquely specifying the classical nonlinear equations of motion and thus the couplings 
between the matter and the gauge fields of the HSS. However, most of the discussion in the rest of the paper does not rely directly on
this construction; readers who want to go directly  to the discussion of the matter sector 
and its representation content from the point of view of the HSS may therefore skip the remainder of this section.

\subsection{The Higher Spin Subalgebra}

As discussed above, we will often, for simplicity restrict to a single complex boson instead of the 
$\mathbb{T}^4$ theory. What this 
means from the point of view of the underlying symmetric product theory is that we are considering single `trace' chiral generators 
in the theory of $N$ complex bosons. We can write down explicit expressions for these generators in terms of the underlying oscillators 
which satisfy 
\be
{}[\alpha^i_m,\bar{\alpha}^j_n] = m \, \delta^{ij}\, \delta_{m,-n} \ , \qquad [\alpha^i_m,\alpha^j_n]=[\bar{\alpha}^i_m,\bar{\alpha}^j_n]=0 \ .
\ee
In the following we will usually suppress the $i,j$ index, since all expressions will be summed over the common index 
$i=1,\ldots,N$.  

Let us first discuss the structure of the higher spin subalgebra ${\rm hs}[1]$. 
The generating fields of this subalgebra were explicitly given in \cite{Bakas:1990ry}, see also 
\cite{Gaberdiel:2013jpa}. 
Since we will only be interested in the wedge algebra ${\rm hs}[1]$ of the ${\cal W}_{\infty}[1]$ algebra, we may work in a 
quasi-primary basis (and not worry about making the fields primary). 
At spin $2$, the relevant modes are just the Virasoro modes,
\be
W^{(2)}_m \equiv L_m = \sum_{n\in\mathbb{Z}} \, : \alpha_{n} \bar{\alpha}_{m-n} :  \ ,
\ee
while the modes of the spin $3$ field are
\be\label{Wmode}
W^{(3)}_m \equiv W_m = 2  \sum_{n\in\mathbb{Z}} (2n-m) \, : \alpha_{n} \bar\alpha_{m-n} :  \ ,
\ee
and for the spin $4$ field we get
\be\label{Umode}
W^{(4)}_m \equiv U_m =  \frac{16}{5} \, \sum_{n\in\mathbb{Z}} \, (m^2-5mn+5 n^2+1)\, : \alpha_{n} \bar\alpha_{m-n} :   \ ,
\ee
etc. [We are using here the conventions of the paper \cite{Gaberdiel:2013jpa}, except that we have changed
the signs of the generators (as well as the commutator of the generators).] Continuing in this manner we can see that 
the modes of the spin $s$ field take the form 
\be
W^{(s)}_m = \sum_{n\in\mathbb{Z}} \, f^{(s)}(m,n) \, : \alpha_{n} \bar\alpha_{m-n} : \ , 
\ee
where $f^{(s)}(m,n)$ is a polynomial in $m$ and $n$ of total degree $s-2$. Because the wedge modes, i.e., the modes
$W^{(s)}_m$ with $|m|\leq s-1$, annihilate the in- and
out-vacuum, the polynomial has the property that 
\be
f^{(s)}(m,n) = 0 \qquad \hbox{if $|m|\leq s-1$ and $n (m-n)>0$.}
\ee
One easily checks that this is indeed the case for the above expressions. Thus all of the wedge modes will be linear combinations of the bilinears 
\be
: \alpha_{n} \, \bar\alpha_{m} : \qquad n \cdot m \leq 0 \ . 
\ee
On any given Fock space state, only finitely many such terms (for a given $m$) have a non-trivial action, and hence
we may think of ${\rm hs}[1]$ as being spanned by the normal ordered generators
\be
E_{r,s} \equiv \alpha_{-r} \bar\alpha_{s} \quad \hbox{and} \quad 
\overline{E}_{r,s} \equiv \bar\alpha_{-r} \alpha_{s} \qquad \hbox{where $r,s\geq 0$.}
\ee
These generators satisfy the commutation relations
\be\label{ELie}
{}[E_{r_1,s_1}, E_{r_2,s_2}] = s_1 \delta_{s_1,r_2} \, E_{r_1,s_2} - r_1 \delta_{r_1,s_2} \, E_{r_2,s_1} \ ,
\ee
and similarly for $\overline{E}_{r,s}$; on the other hand, the mutual commutator $[E_{r_1,s_1},\overline{E}_{r_2,s_2}]=0$
vanishes. If we define 
\be
F_{r,s} \equiv \left\{ \begin{array}{ll}
E_{r,s} \quad & \hbox{if $r,s\geq 0$} \\
\overline{E}_{-r,-s} \quad & \hbox{if $r,s\leq 0$} 
\end{array}
\right.
\ee
then the commutation relations take the compact form
\be
{}[F_{r_1,s_1}, F_{r_2,s_2}] = |s_1| \delta_{s_1,r_2} \, F_{r_1,s_2} - |r_1| \delta_{r_1,s_2} \, F_{r_2,s_1} \ .
\ee
\smallskip

We should note that the subalgebra generated by $E_{r,s}$ with $r,s>0$ (and similarly for $\overline{E}_{r,s}$)
is isomorphic to the Lie algebra $\mathfrak{gl}(\infty)$, i.e., the Lie algebra of infinite dimensional matrices subject to the 
condition that only finitely many matrix entries are non-zero. Let us concentrate, for definiteness, on the generators
$E_{r,s}$ with $r,s>0$, for which it is convenient to introduce the rescaled generators
\be
\hat{E}_{r,s} \equiv \frac{1}{\sqrt{rs}}\, E_{r,s} \ . 
\ee
Then the above Lie bracket (\ref{ELie}) comes from the associative product
\be\label{associative}
\hat{E}_{r_1,s_1} \ast \hat{E}_{r_2,s_2} = \delta_{s_1,r_2} \, \hat{E}_{r_1,s_2} \ ,
\ee
which is nothing but the matrix product of infinite-dimensional matrices in which $\hat{E}_{r,s}$ has a non-zero entry 
(equal to $1$) only for $(r,s)$. We should also note that the corresponding Lie algebra,
i.e., the Lie algebra generated by the $E_{r,s}$ or $\hat{E}_{r,s}$ with $r,s>0$, is already
isomorphic to ${\rm hs}[1]$. To see this, we define the $\mathfrak{sl}(2)$ generators 
\be\label{sl2}
L_n = \sum_{\min(i,i+n)\geq 1}^{\infty}  E_{i,i+n} = 
\sum_{\min(i,i+n)\geq 1}^{\infty} \sqrt{i(i+n)} \, \hat{E}_{i,i+n} \ , \qquad n=0, \pm 1 \ ,
\ee
which one easily checks to satisfy the commutation relations
\be
[L_m,L_n] = (m-n) \, L_{m+n} \ . 
\ee
Using the associative product (\ref{associative}) we can then consider the universal enveloping algebra $U(\mathfrak{sl}(2))$ of this
$\mathfrak{sl}(2)$ algebra which agrees, as a vector space, with the whole of $\mathfrak{gl}(\infty)$.
Finally, we calculate the Casimir of this $\mathfrak{sl}(2)$, and we find
\be
C =  L_0^2 - \frac{1}{2} \bigl( L_1 L_{-1} + L_{-1} L_1 \bigr) = 0  =  \left. \frac{1}{4} \bigl( \lambda^2 - 1 \bigr) \right|_{\lambda=1} \ ,
\ee
i.e., the value of the Casimir agrees with that of ${\rm hs}[1]$. 

\subsubsection{The Differential Algebra Viewpoint}

Another useful way to think about this wedge algebra is in terms of differential operators. Let us consider infinitely
many coordinates $z_i$ and $\bar{z}_i$, where $i=1,2,\ldots$. Then we can identify 
\be\label{i1}
\alpha_{-m} \ \longleftrightarrow \ z_m \qquad \bar\alpha_{-m} \ \longleftrightarrow \ \bar{z}_m \qquad \quad 
(m>0)
\ee
and 
\be\label{i2}
\alpha_{m} \ \longleftrightarrow \ m \, \partial_{z_m} \qquad \bar\alpha_{m} \ \longleftrightarrow \ m \, \partial_{\bar{z}_m} \qquad \quad
(m>0) \ . 
\ee
Note that since $\alpha_0$ and $\bar{\alpha}_0$ are central, they can be thought of as numbers; for the present discussion we 
shall assume that both central elements are zero, although it is not difficult to generalise also to the situation where they
take non-zero values. The wedge algebra ${\rm hs}[1]$ can then be thought of as being spanned by the differential operators
\be\label{bilindif}
z_i \, \partial_{\bar{z}_j} \qquad \hbox{and} \qquad \bar{z}_i \, \partial_{z_j} \ .
\ee
In turn, these operators can be interpreted as the generators of rotations, scalings and shear transformations on this large space.

\subsection{The Stringy Algebra}\label{sec:SA}

This last viewpoint now also suggests how one may extend the  higher spin algebra ${\rm hs}[1]$
to the Higher Spin Square. To see this, recall that the chiral algebra of the symmetric orbifold also contains generators such as 
\be\label{Vgen}
V =  \sum_i \partial \phi^i \, \partial\phi^i \,\, \partial\bar\phi^i\, \partial\bar\phi^i\ , 
\ee
whose mode expansion is simply (again suppressing the $i$-index)
\be\label{Vmod}
V_m = \sum_{p,r,s}\, :\alpha_{p} \, \alpha_{r} \, \bar\alpha_{s} \, \bar\alpha_{m-p-r-s} : \ . 
\ee
The corresponding wedge algebra thus also contains linear combinations of generators of the form 
\be\label{HSSwedge}
: \alpha_{n_1}\cdots \alpha_{n_l} \, \bar\alpha_{m_1} \cdots \bar\alpha_{m_r} :  \ , 
\ee
where at least one of the modes is non-negative, and one of the modes is non-positive (so as to annihilate the
in- and out-vacuum). There exists an interesting subalgebra of `neutral' fields for which $l=r$ --- the field $V$ from above
is in fact the lowest neutral field that is not just bilinear.
We have 
worked out some of the low-lying commutators of the spin $4$ field $V$ with the bilinear higher spin currents 
in Appendix~\ref{app:sa}. 

In terms of the differential algebra viewpoint, the generators of the HSS algebra are
then of the form 
\be
\prod_{p=1}^{m} z_{i_p}^{(\epsilon_p)} \, \prod_{q=1}^{n} \partial_{z_{j_q}^{(\epsilon_q)}}  \ , 
\ee
where $n,m>0$, and we use the convention that 
\be
z_i^{\epsilon} \equiv \left\{ 
\begin{array}{ll}
z_i \quad & \hbox{if $\epsilon=+$} \\
\bar{z}_i \quad & \hbox{if $\epsilon=-$.} 
\end{array}
\right.
\ee
(For the neutral HSS we therefore have as many $\epsilon =+$ as $\epsilon=-$.) 
Thus the HSS algebra is some higher order extension of the diffeomorphism algebra involving 
differential operators of arbitrary order.\footnote{Some aspects of its representation theory were
analysed in \cite{BBF}.}
 We suspect that this viewpoint will eventually be useful in understanding 
the geometric interpretation of this symmetry as well as for constructing the couplings to matter fields.

\section{The Untwisted Sector}

We now begin our investigations into the matter content of string theory at the symmetric product orbifold point. 
From the CFT side,  the untwisted sector of the symmetric orbifold of $\mathbb{T}^4$ is a natural place to start. 
As mentioned in the introduction, we will see that the entire matter sector here can be viewed as the Fock space of 
a single irreducible representation of the HSS --- the minimal representation.  

For concreteness we will be working in the NS-NS sector; obviously, the other sectors can be obtained from this
by spectral flow. In the large $N$ limit the partition function of the symmetric 
orbifold in the untwisted sector takes the form \cite{Dijkgraaf:1996xw}
\be\label{symorb}
q^{\frac{N}{4}} \, \bar{q}^{\frac{N}{4}} \, \, \Z_{\rm U}(q,y,\bar{q},\bar{y}) 
= {\prod_{r,\bar{r}=0}^{\infty}}{}' \, \prod_{l,\bar{l}\in\mathbb{Z}}\, \Bigl( 1 - (-1)^{2r + 2\bar{r}}\, q^r y^l \bar{q}^{\bar{r}} \bar{y}^{\bar{l}} \Bigr)^{-d(r,l) 
d(\bar{r},\bar{l})} \ , 
\ee
where the prime at the product means that the term with $r=\bar{r}=0$ is excluded. (This is the term associated to the vacuum that is not
`multiparticled'.) Furthermore, the $d(r,l)$ are the expansion coefficients
\begin{eqnarray}
\tilde{Z}_{\rm chiral}(q,y)  & = & \prod_{n=1}^{\infty} \frac{\bigl( 1 - y q^{n-1/2}\bigr)^2 \, \bigl( 1 - y^{-1} q^{n-1/2} \bigr)^2}{(1-q^n)^4} 
\ = \ \sum_{r=0,l} d(r,l) \, q^r y^l \nonumber \\ 
& = & 1 - 2(y+y^{-1}) q^{1/2} + (y^2 + 8 + y^{-2}) q^1 - 12 (y + y^{-1}) q^{3/2}   \nonumber \\
& & \quad + (8y^2 + 39 + 8 y^{-2})\, q^2 + \cdots \ ,
\end{eqnarray}
i.e., of the character (with insertion of $(-1)^F$) of a single $\mathbb{T}^4$. Note that the $d(r,l)$ have alternating signs, i.e.,
$d(r,l)=(-1)^{2r}\tilde{d}(r,l)$ with $\tilde{d}(r,l)$ being positive integers. 
Because of these alternating signs of $d(r,l)$ and the presence of the $(-1)^{2r}$ factors in (\ref{symorb}), $\Z_{\rm U}$ is the full 
generating function of all states in the orbifold CFT (and not an index).  

We can decompose both the left- and right-moving contributions to the partition function in terms of representations of the
symmetric group, and then the answer takes the form 
\be\label{decom}
\Z_{\rm U}(q,y,\bar{q},\bar{y})  = \sum_R |\chi_R(q,y)|^2 \ ,
\ee
where the sum runs over all representations of $S_N$, the symmetric group (i.e., all Young diagrams with $N$ boxes). Each, say, 
left moving term $\chi_R(q,y)$ is the total contribution of all states that transform in the representation $R$ of $S_N$. Here we have 
also used that the singlet condition with respect to $S_N$ implies that the only combinations that appear involve the same 
representation of $S_N$ for the left- and right-movers (and that each such term appears with multiplicity one). 


\subsection{The Minimal Representation and its Tensor Powers}

While the above decomposition is true for the symmetric orbifold for any finite $N$, we will see a somewhat stronger 
manifestation of the multiparticling at large $N$ as follows. We will show below that 
\be\label{2.14a}
q^{\frac{N}{4}} \, \bar{q}^{\frac{N}{4}} \, \, \Z_{\rm U}(q,y,\bar{q},\bar{y}) = 
|{Z}_{\rm vac}|^2 \cdot \sum_{R} |\Phi_{R}^{({\rm wedge})}(q,y)|^2 \ .
\ee
Here ${Z}_{\rm vac}$ is the vacuum character associated to the Higher Spin Square (HSS), see eq.~(\ref{zvac}) below,
and $\Phi_{R}^{({\rm wedge})}(q,y)$ are the  multiparticle characters of the minimal representation of the HSS.
(In order to avoid confusion, we shall always denote the characters of the Higher Spin Square by capital greek letters,
whereas those of the original higher spin algebra will be denoted by lower case greek letters.)
Note that the sum over $R$ here is over {\it all} Young diagrams (i.e., not just those with $N$ boxes),  and that it describes
the symmetrisation or anti-symmetrisation of the minimal representation. Thus (\ref{2.14a}) can indeed be viewed as the higher spin like 
decomposition of the partition function in terms of the minimal representation of the HSS, as well as its tensor powers.\footnote{A direct 
passage from 
(\ref{decom}) to (\ref{2.14a}) is not obvious at all since the Young diagrams that appear in (\ref{decom}) have $N$ boxes, while the Young
diagrams in (\ref{2.14a}) have an arbitrary number of boxes. There is, however, a natural correspondence between the two sets of Young 
diagrams in the large $N$ limit: every Young diagram $R'$ in (\ref{2.14a}) can be `completed' to one having $N$ boxes by 
adding a first row with sufficiently 
many boxes. So the empty Young diagram $R'$ corresponds to the
Young diagram $R$ with one row of $N$ boxes, the single box for $R'$ corresponds to the Young diagram $R$ with
$N-1$ boxes in the first row, and one in the second, etc. } Let us spend a moment on what this means before we actually demonstrate the result.

Recall that in the conventional Vasiliev theory, the minimal representation
describes the degrees of freedom of a (massive) scalar field multiplet \cite{Gaberdiel:2013vva}, and that the perturbative part of the higher spin theory consists of the multi-particle contributions of this representation \cite{Gaberdiel:2014cha}. Thus, by showing that the terms in (\ref{decom}) can be interpreted as describing the symmetrised and anti-symmetrised contributions of the minimal representation, we see that the untwisted sector of the 
symmetric orbifold has the same structure as the perturbative part of a higher spin theory --- 
the only difference being that the corresponding higher spin algebra is not vector-like,
but rather the Higher Spin Square (i.e., the stringy extension of the higher spin algebra) 
of \cite{Gaberdiel:2015mra}. In particular, we would expect the coupling of this minimal representation to the HSS gauge 
fields to be like that in the Vasiliev theory except that the higher spin algebra is replaced by the HSS algebra. It would be 
very interesting to work this out explicitly.

In order to show that we can indeed write $\Z_{\rm U}$ in this manner, 
we first need to recall the structure of the `minimal' representation
for the HSS. It was shown in \cite{Gaberdiel:2014cha}, see Section~4.3 of that paper, 
that the first non-trivial representation besides 
the extended vacuum representation (that is associated with the trivial representation of the symmetric group) 
has the wedge character 
\begin{eqnarray}\label{minHSSwedge}
\Phi^{(\rm wedge)}_1(q,y) & =  & Z_{\rm chiral}(q,y) - 1 \nonumber \\
& = & (2y + 2y^{-1}) q^{1/2} + (y^2 + 8 +  y^{-2}) q + (12 y + 12 y^{-1} ) q^{3/2}  
 \nonumber \\
& & \quad + (8y^2 + 39 + 8 y^{-2}) q^2 + \cdots \ , \nonumber \\
& = & \sum_{r>0,\, l} \tilde{d}(r,l)\, q^r y^l  \ ,
\end{eqnarray}
where $Z_{\rm chiral}(q,y)$ is now the character (without insertion of $(-1)^F$) of a single 
$\mathbb{T}^4$
\be
Z_{\rm chiral}(q,y) = \prod_{n=1}^{\infty} \frac{\bigl( 1 + y q^{n-1/2}\bigr)^2 \, \bigl( 1 + y^{-1} q^{n-1/2} \bigr)^2}{(1-q^n)^4} \ .
\ee
Note that
the full character associated to this wedge representation is then the product $\Phi^{(\rm wedge)}_1 \cdot {Z}_{\rm vac}$, where 
${Z}_{\rm vac}$ is the extended vacuum character associated to the Higher Spin Square, see eq.~(4.22) of \cite{Gaberdiel:2014cha}.
The representation $\Phi^{(\rm wedge)}_1$ is the representation associated to the `standard' $(N-1)$-dimensional representation
of $S_N$, whose Young diagram has $(N-1)$ boxes in the first row, and one box in the second (see footnote 1);
it is the representation that will play the role of the `minimal representation' for the Higher Spin Square.
\smallskip

It is instructive to compare the minimal representation of the HSS with the minimal representation of the 
(vertical) ${\cal N}=4$ higher spin algebra, which has the character (see (B.6) of 
\cite{Gaberdiel:2014cha})
\be
\phi^{(\rm wedge)}_1(q,y) = \frac{q^{1/2}}{(1-q)}\,  \bigl(y+y^{-1}  +  2 q^{1/2}\bigr)\ .
\ee
This is very much smaller than the character $\Phi^{(\rm wedge)}_1$ of the HSS. We can identify 
$\phi^{(\rm wedge)}_1$ with the wedge part of the ${\cal N}=4$ coset representation $(0; {\rm f})$ (see \cite{Gaberdiel:2013vva}). 
The minimal representation of the HSS thus contains the minimal representation of the HS algebra but 
also an infinite number of other coset representations. In fact, we have the decomposition, already from eqs.~(2.12) and (2.13) of \cite{Gaberdiel:2015mra},
\be
\Phi^{(\rm wedge)}_1(q,y)  = \sum_{m,n=0}^{\infty\, '}
\chi^{({\rm wedge})}_{(0;[m,0,\ldots,0,n])}(q,y) \ , 
\ee
where the prime indicates that the term with $(m,n)=(0,0)$ is excluded. 
 

Later, when we will be dealing with the simpler case of a complex boson, the analogous character of the HSS
minimal representation is 
\be\label{HSSmin}
\Phi^{(\rm wedge)}_{1({\rm bos})}(q,y) =  \prod_{n=1}^{\infty} \frac{1}{(1- yq^n) (1-y^{-1} q^n)} -1 \ .
\ee
This is again to be contrasted with the minimal representation of the bosonic ${\rm hs}[1]$ algebra which 
equals \cite{Gaberdiel:2010pz,Gaberdiel:2011zw},
\be\label{bosmin}
\phi^{(\rm wedge)}_{1({\rm bos})}(q,y) = \frac{q}{1-q} \ .
\ee
\smallskip

After this brief interlude, let us 
return to eq.~(\ref{decom}). From the point of view of the wedge algebra, the other representations that appear in (\ref{decom}) 
should be associated to the symmetric powers of the minimal HSS representation, i.e., their wedge characters should be 
the symmetrised and/or antisymmetrised powers of $\Phi^{(\rm wedge)}_1$. Because we are dealing with supersymmetric
theories, one has to be a little careful with how the symmetrisation and anti-symmetrisation is performed. For example,
for the second power, the symmetrised and anti-symmetrised powers then have the character
\be\label{3.10}
\Phi_{{\rm sym}}^{(\rm wedge)}(q,y) = \frac{1}{2} \Bigl( \Phi^{(\rm wedge)}_1(q,y)^2 + 
{\Phi}^{(\rm wedge),-}_1(q^2,y^2) \Bigr) \ , 
\ee
and
\be\label{3.11}
\Phi_{{\rm anti-sym}}^{(\rm wedge)}(q,y) = \frac{1}{2} \Bigl( \Phi^{(\rm wedge)}_1(q,y)^2 -
{\Phi}^{(\rm wedge),-}_1(q^2,y^2) \Bigr) \ ,
\ee
where ${\Phi}^{(\rm wedge),-}_1$ is the wedge character of $\Phi_1$ with the insertion of $(-1)^F$
\be\label{minHSSwedget}
\Phi^{(\rm wedge), -}_1(q,y) = \tilde{Z}_{\rm chiral}(q,y) - 1 = \sum_{r>0, l} d(r,l)q^r y^l  \ .
\ee 

For the case at hand, it is not hard to work out the lowest order terms in $\Phi_{{\rm sym}}^{(\rm wedge)}$ and 
$\Phi_{{\rm anti-sym}}^{(\rm wedge)}$, and one
finds
\begin{eqnarray}\label{twobox}
\Phi_{{\rm sym}}^{(\rm wedge)} (q,y) & = & (y^2 + 4 + y^{-2}) q^1 + (2 y^3 + 18 y + 18y^{-1} + 2 y^{-3}) q^{3/2} \nonumber \\
& & \quad + (y^4 + 32 y^2 + 85 + 32 y^{-2} + y^{-4}) q^2 + \cdots 
\\[4pt]
\Phi_{{\rm anti-sym}}^{(\rm wedge)} (q,y) & = &  (3y^2 + 4 + 3y^{-2}) q^1 + (2 y^3 + 18 y + 18y^{-1} + 2 y^{-3}) q^{3/2} \nonumber \\
& & \quad + (32 y^2 + 77 + 32 y^{-2} ) q^2 + \cdots \ .
\end{eqnarray}
Incidentally, in terms of the ${\cal W}_{\infty}$ representations, this is compatible with 
\begin{eqnarray}
{Z}_{\rm vac}\cdot \Phi_{{\rm sym}}  & = & \chi_{(0;[2,0,\ldots,0])} + \chi_{(0;[0,\ldots,0,2])} + \chi_{(0;[1,0,\ldots,0,1])} + \cdots \\
{Z}_{\rm vac}\cdot \Phi_{{\rm anti-sym}} & = & \chi_{(0;[0,1,0,\ldots,0])} + \chi_{(0;[0,\ldots,0,1,0])} + \chi_{(0;[1,0,\ldots,0,1])} 
+ \cdots \ .
\end{eqnarray}

\noindent For the case with three boxes, the relevant combinations are 
\be\label{3box1}
\Phi_{{\tiny \yng(3)}}^{(\rm wedge)}(q,y) = \frac{1}{6} \Bigl( \Phi^{(\rm wedge)}_1(q,y)^3 
+ 3 {\Phi}^{(\rm wedge),-}_1(q^2,y^2)  \Phi^{(\rm wedge)}_1(q,y)
+ 2  \Phi^{(\rm wedge)}_1(q^3,y^3) \Bigr) \ , 
\ee
\be\label{3box2}
\Phi_{{\tiny \yng(1,1,1)}}^{(\rm wedge)}(q,y) = \frac{1}{6} \Bigl( \Phi^{(\rm wedge)}_1(q,y)^3 
- 3 {\Phi}^{(\rm wedge),-}_1(q^2,y^2)  \Phi^{(\rm wedge)}_1(q,y)
+ 2  \Phi^{(\rm wedge)}_1(q^3,y^3) \Bigr) \ , 
\ee
as well as 
\be\label{3box3}
\Phi_{{\tiny \yng(2,1)}}^{(\rm wedge)}(q,y) = \frac{1}{3} \Bigl( \Phi^{(\rm wedge)}_1(q,y)^3 -  \Phi^{(\rm wedge)}_1(q^3,y^3) \Bigr) \ . 
\ee
Then the leading terms of the wedge characters are
\be
\Phi_{{\tiny \yng(3)}}^{(\rm wedge)}(q,y) = (4 y^3 + 6 y + 6 y^{-1} + 4 y^{-3}) q^{3/2} + 
(3y^4 + 28 y^2 + 38 + 28 y^{-2} + 3 y^{-4}) q^2 + \cdots \ , 
\ee
\be
\Phi_{{\tiny \yng(1,1,1)}}^{(\rm wedge)}(q,y) = (2y + 2y^{-1}) q^{3/2} + 
(y^4+ 12 y^2 + 34+ 12 y^{-2} + y^{-4}) q^2 + \cdots \ , 
\ee
as well as 
\be
\Phi_{{\tiny \yng(2,1)}}^{(\rm wedge)}(q,y) = (2y^3 + 8y + 8y^{-1} + 2 y^{-3}) q^{3/2} 
+ (4y^4 + 40 y^2 + 72 + 40 y^{-2} + 4 y^{-4}) q^2 + \cdots \ . 
\ee
The generalisation to arbitrary symmetrisation (or anti-symmetrisation) is given by the 
so-called Frobenius formula 
\be\label{Frob}
\Phi_R^{({\rm wedge})}(q,y) = \sum_{\vec{k}}\frac{1}{z_{\vec{k}}}\, \chi_R(C(\vec{k}))\, \Phi^{({\rm wedge})}_{\vec{k}}(q,y) \ ,
\ee
where $R$ is a Young diagram, and $\chi_R(C(\vec{k}))$ is the character in the representation $R$ 
of the group element in the conjugacy class labelled by the vector $\vec{k}= \{ k_i, k_2, \ldots \}$, with the convention that 
the relevant permutation has a cycle structure with $k_j$ cycles of length $j$. Furthermore, 
\be
z_{\vec{k}} = \prod_j (k_j)! \prod_j j^{k_j} \ ,
\ee
and 
\be\label{chieps}
\Phi^{({\rm wedge})}_{\vec{k}}(q,y) = \prod_{j} \Bigl[ \Phi_1^{({\rm wedge}), \epsilon_j}(q^j,y^j) \Bigr]^{k_j} \ , 
\ee
where $\epsilon_j = (-1)^{j+1}$ denotes whether the character is the standard character (if $\epsilon=+1$),
or the character with the insertion of $(-1)^F$ (if $\epsilon=-1$).
It is easy to see that (\ref{Frob}) reduces to eqs.~(\ref{3.10}), (\ref{3.11}), (\ref{3box1}), (\ref{3box2}), and (\ref{3box3}) for the 
corresponding choices of Young diagram $R$. 

The claim above is now equivalent to the statement that, in general, 
eq.~(\ref{symorb}) can be written as 
\be\label{2.14}
q^{\frac{N}{4}} \, \bar{q}^{\frac{N}{4}} \, \, \Z_{\rm U}(q,y,\bar{q},\bar{y}) = 
|{Z}_{\rm vac}|^2 \cdot \sum_{R} |\Phi_{R}^{({\rm wedge})}(q,y)|^2 \ , 
\ee
where $\Phi_{R}^{({\rm wedge})}(q,y)$ is defined by eq.~(\ref{Frob}).  In the next subsection, we give a proof of this relation. 
As a reality check, we have also verified this explicitly for the first few terms, i.e., up to order $q^{3/2} \bar{q}^2$ and $q^2\bar{q}^{3/2}$.

\subsection{Proof of Multi-particling of Minimal Representation}

In this subsection we prove eq.~(\ref{2.14}) to arbitrary order, i.e., show that 
the entire untwisted partition function in eq.~(\ref{symorb}) can be described by multi-particling the minimal 
representation. One way to see this is to rewrite eq.~(\ref{symorb}) as 
\begin{eqnarray}\label{multmin}
q^{\frac{N}{4}} \, \bar{q}^{\frac{N}{4}} & & \Z_{\rm U}(q,y,\bar{q},\bar{y}) 
=  \prod_{r,\bar{r}=0}^{\infty}{}' \, \prod_{l,\bar{l}\in\mathbb{Z}}\, 
\Bigl( 1 - (-1)^{2r + 2\bar{r}}\, q^r y^l \bar{q}^{\bar{r}} \bar{y}^{\bar{l}} \Bigr)^{-d(r,l) d(\bar{r},\bar{l})}  \\
&&  =    \exp{\Bigl(-\sum_{r,\bar{r}=0}^{\infty}{}'\sum_{l,\bar{l}\in\mathbb{Z}} \, d(r,l) \,
d(\bar{r},\bar{l})} \ln{\Bigl( 1 - (-1)^{2r + 2\bar{r}}\, q^r y^l \bar{q}^{\bar{r}} \bar{y}^{\bar{l}} \Bigr)}\Bigr)  \cr 
&&= \exp{\Bigl(\sum_{r,\bar{r}=0}^{ \infty}{}'\sum_{l,\bar{l}\in\mathbb{Z}}\, d(r,l) \,
d(\bar{r},\bar{l})} \sum_{k=1}^{\infty}\frac{1}{k}(-1)^{(2r + 2\bar{r})k}\, q^{kr} y^{kl} \bar{q}^{k\bar{r}} \bar{y}^{k\bar{l}} \Bigr)  \cr 
&& =  \exp{\Bigl(\sum_{r,\bar{r}=0}^{\infty}{}'\sum_{l,\bar{l}\in\mathbb{Z}}\, \tilde{d}(r,l) \,
\tilde{d}(\bar{r},\bar{l})} \sum_{k=1}^{\infty}\frac{1}{k}(-1)^{(2r + 2\bar{r})(k+1)}\, q^{kr} y^{kl} \bar{q}^{k\bar{r}} \bar{y}^{k\bar{l}} \Bigr)  \cr 
&& =  \exp{\Bigl(\sum_{k=1}^{\infty}\frac{1}{k}\Bigl[\bigl(1+{\Phi}^{({\rm wedge}),\epsilon_k}_1(q^k,y^k)\bigr)
\bigl(1+{\Phi}^{({\rm wedge}),\epsilon_k}_1(\bar{q}^k,\bar{y}^k)\bigr) -1\Bigr] \Bigr)} \ , \nonumber 
\end{eqnarray}
where we have used eq.~(\ref{minHSSwedge}) as well as $d(r,l) = (-1)^{2r}\, \tilde{d}(r,l)$ and 
$\epsilon_k=(-1)^{k+1}$, see the definition given below eq.~(\ref{chieps}). Multiplying out the brackets in the exponent and using
that 
\begin{eqnarray}
\sum_{k=1}^\infty \frac{1}{k} {\Phi}^{({\rm wedge}),\epsilon_k}_1(q^k,y^k) & =  & 
\sum_{r>0,l} \tilde{d}(r,l) (-1)^{2r} \sum_{k=1}^{\infty} \frac{1}{k} q^{rk} y^{lk} (-1)^{2rk}  \nonumber \\
& = & - \sum_{r>0,l} \, d(r,l) \, \ln{\Bigl( 1 - (-1)^{2r}\, q^r y^l \Bigr)} \ ,
\end{eqnarray}
and similarly for the right-movers, the last expression can now be rewritten as 
\begin{eqnarray}\label{multmin1}
& = &\Bigl|\prod_{r=0}^{\infty} \, \prod_{l \in\mathbb{Z}}\, \Bigl( 1 - (-1)^{2r}\, q^r y^l  \Bigr)^{-d(r,l)}\Bigr|^2 
\exp{\Bigl(\sum_{k=1}^{\infty}\frac{1}{k}
{\Phi}^{({\rm wedge}),\epsilon_k}_1(q^k,y^k) {\Phi}^{({\rm wedge}),\epsilon_k}_1(\bar{q}^k,\bar{y}^k) \Bigr)} \cr
& =& | q^{\frac{N}{4}} {Z}_{\rm vac}|^2 \, \exp{\Bigl(\sum_{k=1}^{\infty}\frac{1}{k}\, 
{\Phi}^{({\rm wedge}),\epsilon_k}_1(q^k,y^k)\,  {\Phi}^{({\rm wedge}),\epsilon_k}_1(\bar{q}^k,\bar{y}^k) \Bigr)}\ ,
\end{eqnarray}
where we have used that (see eq.~(2.6) of \cite{Gaberdiel:2015mra}) 
\be\label{zvac}
q^{\frac{N}{4}} {Z}_{\rm vac} = \prod_{r=0}^{\infty} \, \prod_{l \in\mathbb{Z}}\, \Bigl( 1 - (-1)^{2r}\, q^r y^l  \Bigr)^{-d(r,l)} \ .
\ee
This now shows the factorisation into the vacuum character, and an exponentiation of the 
wedge character of the minimal representation. Indeed the latter factor is precisely of the form such that it can be 
expanded in terms of the representations that arise in the tensor product of finite powers of the minimal representation. 

To see this we use the general group theoretic identities (see, e.g., eq.~(3.2) and below of \cite{Dutta:2007ws})
\begin{eqnarray}\label{grpid}
\sum_R\chi_R(U)\chi_R(V) & = & 1+\sum_{\vec{k}}\frac{1}{z_{\vec{k}}}\Xi_{\vec{k}}(U)\, \Xi_{\vec{k}}(V) \ , \cr
& = & \exp{\Bigl[\sum_{k=1}^{\infty}\frac{1}{k} {\rm Tr}(U^k){\rm Tr}(V^k) \Bigr]} \ . 
\end{eqnarray}
In the first line, the LHS consists of a sum over all $R$, i.e., all Young diagrams with an arbitrary  number of boxes, while
the functions on the RHS are 
\be
\Xi_{\vec{k}}(U) =  \prod_{i} {\rm Tr}(U^{k_i}) \ .
\ee
Eq.~(\ref{grpid}) can be proven by using the Frobenius relation (\ref{Frob}) and the orthogonality of characters of the symmetric group. 
The second line is obtained by expanding out the exponent of the right-hand-side, and gathering terms  with the same $\vec{k}$. 
In the current context this then leads to the identity 
\be
\exp{\Bigl(\sum_{k=1}^{\infty}\frac{1}{k}{\Phi}^{(\rm wedge),\epsilon_k}_1(q^k,y^k)
{\Phi}^{(\rm wedge),\epsilon_k}_1(\bar{q}^k,\bar{y}^k) \Bigr)} = 
\sum_R {\Phi}^{(\rm wedge)}_R(q, y) \, {\Phi}^{(\rm wedge)}_R (\bar{q}, \bar{y}) \ .
\ee
Together with eq.~(\ref{multmin1}) this then proves (\ref{2.14}).

\section{The Twisted Sector from the HS viewpoint}

In the previous section we have seen that the untwisted sector of the symmetric orbifold can be interpreted 
as the perturbative part of the higher spin theory associated to the Higher Spin Square (HSS), where we add a 
massive scalar field, as well as its HSS descendants 
to the higher spin gauge field degrees of freedom. Our next aim is to understand the structure of the twisted sector from this viewpoint.
In this section we shall, as a first step, concentrate on describing the twisted sector from the viewpoint of the HS algebra (rather than
the HSS); we shall return to what can be said from the viewpoint of the HSS in Sec.~\ref{sec:twisHSS}. 

For simplicity, let us first study the bosonic toy model, i.e., the symmetric orbifold of $N$ complex bosons; thus
we will analyse the twisted sector from the viewpoint of the ${\rm hs}[1]$ algebra.
Since we are interested in the single-particle states, we shall only consider twisted sectors associated to 
a single cycle of length $m$, say. This twisted sector is then generated by the modes
\be
\alpha^{(r)}_{n+\frac{r}{m}} \ , \qquad \bar\alpha^{(r)}_{n-\frac{r}{m}} \ , \qquad r=0,\ldots, m-1 \ , 
\ee
for which the commutation relations take the form
\be
{}[ \alpha^{(r)}_{t}, \bar\alpha^{(s)}_{u}] = \delta^{rs} \, t \,  \delta_{t,-u} \ .
\ee
Note that the modes associated to the different twists $r$ commute with one another. We will therefore study, to begin with, 
the contributions that come from one such twist $\nu\in(0,1)$, i.e., the vector space that is generated by the modes 
\be
\alpha_{n+\nu} \ , \qquad \bar\alpha_{n-\nu} \ , \qquad \ \ n\in\mathbb{Z} \ .
\ee
To describe a given twisted sector of the symmetric orbifold we will then, in the end, have to add the contributions 
coming from the different twists $\nu =\frac{r}{m}$ together. For the bilinear generators associated to 
${\rm hs}[1]$ this is straightforward since they will always be sums of modes from the same $\nu$-twisted
sector; for the stringy higher order generators there will also be in general mixed terms, and thus the structure of the actual 
$m$-cycle twisted sector representation will be somewhat more complicated. (As a consequence we are, perforce, somewhat less specific in that case.)

For the first few ${\rm hs}[1]$ generators it is not difficult to work out what the contribution to the low-lying wedge modes are, 
and one finds 
\be\label{Ltwist}
L_m = \sum_{r\in\mathbb{Z}+\nu} \, : \alpha_{r} \bar{\alpha}_{m-r} :   \ + \ \frac{1}{2}\, \nu \, (1-\nu) \, \delta_{m,0} \ ,
\ee
while the modes of the spin $3$ field are
\be\label{Wtwist}
W_m = 2  \sum_{r\in\mathbb{Z}+\nu} (2r-m) \, : \alpha_{r} \bar\alpha_{m-r} :  
+ \ \frac{4h}{3} \, (2\nu-1) \, \delta_{m,0}  \ ,
\ee
where $h=\frac{1}{2}\, \nu\, (1-\nu)$, and for the spin $4$ field we get
\be\label{Utwist}
U_m =  \frac{16}{5} \, \sum_{r\in\mathbb{Z}+\nu} \, (m^2-5mr+5 r^2+1)\, : \alpha_{r} \bar\alpha_{m-r} :   
+ \ \frac{16}{5} \frac{h}{2} \, (5 \nu^2 -5 \nu +2 ) \, \delta_{m,0}  \ ,
\ee
etc.  Here, the normal ordering corrections were determined by the condition that these modes have the correct
commutation relations with the Virasoro modes, i.e., that the $L_m$ form a Virasoro algebra, and that 
\be
{} [L_m,W_n] =  (2m-n) W_{m+n} \ , \qquad [L_m,U_n] =  (3m-n) U_{m+n}   + \tfrac{32}{5}\, m (m^2-1)\, L_{m+n}\ . 
\ee
It is maybe worth pointing out that the zero mode shift for the odd spin $s$ fields always contains a factor
of $(2\nu -1)$, i.e., vanishes for $\nu=\frac{1}{2}$.  It is not hard to work out these shifts for all spin fields, but
the general formula does not seem to be very illuminating.

\subsection{The structure of the twisted representation}

In order to understand the structure of the twisted representation, it is now useful to work out 
the various null-vectors it possesses. We shall only consider the action of the wedge modes, i.e., ignore
the multiparticle descendants. First of all, since $0<\nu<1$, all $(-1)$-modes are proportional to 
\be\label{level1}
W^{(s)}_{-1} \phi \, \sim \, \alpha_{-1+\nu} \bar\alpha_{-\nu} \, \phi   \ , 
\ee
and hence must be proportional to $L_{-1}\phi$. (Here $\phi$ is the ground state of the twisted sector, i.e., it is 
annihilated by all positive modes.) The proportionality constant turns out to be 
\be\label{WsN1}
W^{(s)}_{-1} \phi = \frac{ s w^{(s)}}{ 2h} \,  L_{-1}\, \phi \ ,
\ee
since the difference must be annihilated by $L_1$,  where $w^{(s)}$ is the $W^{(s)}_0$ eigenvalue of $\phi$. For example,
for the case of $s=3$, we get from (\ref{Ltwist}) and (\ref{Wtwist}), 
\be \label{NW1}
\frac{3 w^{(3)}}{2h} = 2 (2\nu-1) \ , \qquad \hbox{i.e.} \quad 
W_{-1} \phi = 2 (2\nu-1) L_{-1} \phi \ . 
\ee
Similarly, one finds from the explicit mode expansion of the spin $4$ field  (\ref{Utwist}) the relation
\be\label{NU1}
U_{-1} \, \phi = \frac{16}{5} (5\nu^2-5\nu+2) L_{-1} \, \phi \ , 
\ee
and similarly for the higher spin modes.

At level $2$, any $(-2)$ mode will be equal to a linear combination of the states 
\be\label{level1phi}
W^{(s)}_{-2} \phi \, \sim \, a\, \alpha_{-2+\nu} \bar\alpha_{-\nu} \, \phi  + b\,   \alpha_{-1+\nu} \bar\alpha_{-1-\nu} \, \phi  \ .
\ee
In addition, there is one quartic mode that contributes at level $2$ and that appears, e.g., in the term $L_{-1}^2\,\phi$,
namely
\be
\alpha_{-1+\nu} \alpha_{-1+\nu} \bar\alpha_{-\nu} \bar\alpha_{-\nu} \, \phi \ .
\ee
Thus the space of states at level $2$ is three-dimensional, and we can take it to be spanned by 
\be\label{basis2}
W_{-2}\, \phi \ , \qquad U_{-2}\, \phi \ ,  \qquad L_{-1}^2\, \phi \ .
\ee
For example, we must be able to express $W^{(5)}_{-2} \, \phi$ in terms of these vectors, and one
finds upon an explicit calculation that 
\be\label{level2N}
 W^{(5)}_{-2}\, \phi \equiv X_{-2} \, \phi = -\bigl(16 \nu^2-16 \nu-\frac{32}{7} \bigr) W_{-2} \, \phi + (8\nu-4) \, U_{-2} \, \phi  \ ,
\ee
where $W^{(5)}\equiv X$ is defined in \cite{Gaberdiel:2013jpa}, see also eq.~(\ref{WUcom}).

We can also predict the number of states at higher level. Indeed, the twisted character has the form
\be\label{chinu}
\chi_{[\nu]}(q,y) = q^h\, \prod_{n=1}^{\infty}  (1 - y q^{n-\nu})^{-1} (1 - y^{-1} q^{n-1+\nu})^{-1} \ . 
\ee
[Here $y$ keeps track of whether the mode is of the form $\alpha_{n+\nu}$ or of the form $\bar{\alpha}_{n-\nu}$.] The 
wedge descendants of the highest weight state are then counted by the neutral sector
\be\label{charnu}
\chi_{[\nu]}^{(+)}(q) =  \left. \chi_{[\nu]}(q,y)  \right|_{y^{0}} = q^h \Bigl( 1 + q + 3 q^2 + 6 q^3 + 12 q^4 + \cdots \Bigr) \ .
\ee
This reproduces, in particular, what we have seen above at level $1$ and $2$. 

\subsection{The twisted representation as a special level-one representation}\label{sec:4.2}

The above $\nu$-twisted representation has very few states at low levels --- in particular, it has only
a single state at level one, see eq.~(\ref{charnu}). On the other hand, it is much larger than the
minimal representation of ${\rm hs}[1]$, see eq.~(\ref{bosmin}), since it has a Cardy-like (exponential) growth. This is 
to be contrasted with the representations that appear in finitely many tensor powers of the
minimal representation --- these are the representations that account for the perturbative higher 
spin spectrum --- whose characters only grow polynomially. Thus these $\nu$-twisted representations
appear to be `non-perturbative' from the usual higher spin viewpoint \cite{Castro:2011iw, Gaberdiel:2012ku, Perlmutter:2012ds}; this is also in agreement with
the fact that, in the coset description, they correspond to representations with large (order $k$)
Dynkin labels, see Sec.~\ref{sec:nearmincos}.

The minimal representation of ${\rm hs}[1]$ is completely characterised by the fact that its wedge 
character has one state at level one and one at level two (see Appendix B of \cite{Gaberdiel:2012ku}). 
We could thus similarly try and characterise the $\nu$-twisted representation as the one with a single state at level 
one and at least two states at level two (so as not to be minimal); in the following we shall refer
to these representations as `level-one representations'. However, as is shown in Appendix~\ref{A.1}, 
these conditions by themselves  generically give rise to a very large representation  whose character 
grows as the MacMahon function, see eq.~(\ref{MM}), and that is therefore much bigger than 
eq.~(\ref{chinu}). In fact,  level-one representations, i.e., representations with a single state at level one,
 are characterised  by two quantum numbers, the conformal weight $h$ and the eigenvalue $w^{(3)}$ of $W^{(3)}_0$, 
while for the $\nu$-twisted 
sector representations there is a relation between these two quantum numbers, i.e.,
\be\label{twisquan}
h = \frac{1}{2}\, \nu \, (1-\nu) \ , \qquad w^{(3)} = \frac{2}{3}\, \nu \, (1-\nu) \, (2\nu -1) \ .
\ee
(The fact that there is a two-parameter family of these level one representations also ties up with 
known facts about degenerate representations of cosets, see Sec.~\ref{sec:nearmincos}.) Thus 
the $\nu$-twisted sector representation is a special level-one representation. The first difference
between the number of states of a generic level-one representation and the $\nu$-twisted representation
appears at level $4$, where a generic level-one representation has one more state 
(namely $13$) than the representation that describes the $\nu$-twisted sector; this will be 
explained in Sec.~\ref{sec:twisnm}.

\subsection{The Superconformal Generalisation}

Let us close this section with some comments about the supsersymmetric generalisation of these 
considerations.

For the case with ${\cal N}=2$ superconformal symmetry, the relevant symmetric orbifold is that of the 
${\cal N}=2$ superconformal $\mathbb{T}^2$ theory, consisting of a complex boson and fermion. 
The minimal representation of the supersymmetric higher spin algebra ${\rm shs}[1]$ 
has wedge character equal to, see e.g., \cite{Candu:2012tr}
\be
\chi_{\rm min}^{({\rm wedge}),\, {\cal N}=2} (q) = q^{1/2}\, \frac{(1+q^{1/2})}{(1-q)} \ .
\ee
The corresponding minimal representation of the HSS algebra, on the other hand, has the character
\be
\Upsilon_{\rm min}^{({\rm wedge}),\, {\cal N}=2} (q) = \left. \prod_{n=1}^{\infty} \frac{\bigl(1+y q^{n-\frac{1}{2}}\bigr)\bigl(1+y^{-1} q^{n-\frac{1}{2}}\bigr)}{\bigl(1 - y q^n\bigr) \bigl(1 - y^{-1} q^n\bigr)} \right|_{y^{\pm 1}}  \ ,
\ee
while the character of the $\nu$-twisted sector can be obtained from the analogue of (\ref{charnu}), i.e., 
\begin{eqnarray}\label{twisN2}
\chi_{\rm tw}^{{\cal N}=2} (q,y)  |_{y^0} & = & \left. \prod_{n=1}^{\infty} \frac{\bigl( 1 + y q^{n-\frac{1}{2} -\nu} \bigr)  \, 
\bigl( 1 + y^{-1} q^{n-\frac{1}{2} +\nu} \bigr)}
{\bigl( 1 - y q^{n-\nu} \bigr) \, \bigl( 1 - y^{-1} q^{n-1 -\nu} \bigr)} \right|_{y^0} \nonumber \\
& = & 1 + q^{1/2} + 2 q^1 + 4 q^{3/2} + 7 q^2 + 11 q^{5/2} + 17 q^3 + \cdots \ . 
\end{eqnarray}
As a consistency check one finds that this agrees precisely with the wedge character of the coset representation
\be
\Bigl([m,0,\ldots,0];[m,0,\ldots,0],-m\Bigr) \ ,
\ee
see \cite{Gaberdiel:2014vca,Fredenhagen:2012bw}.\footnote{We thank Maximilian Kelm  for checking this with Mathematica.}

The character (\ref{twisN2}) has only a single state at level $h=1/2$, and hence the underlying representation
is chiral primary, i.e., the ground 
state is annihilated by $G^{-}_{-1/2}\phi=0$, say. (Obviously, there is a choice here about whether $G^+_{-1/2}$
or $G^{-}_{-1/2}$ annihilate the ground state, and the conjugate representation to the above 
will satisfy $G^{+}_{-1/2}\phi=0$.) Furthermore, the same will be true
for any of the fermionic spin $s$ supercharges, i.e., we have the relations
\be\label{minus12}
W^{s-}_{-1/2} \, \phi = 0 
\ee
for all $s=2,3,\ldots$. Here we are using the notation that the spin $s$ multiplet consists of the generators
(see, e.g., \cite{Candu:2012tr})
\be
W^{s0} \ , \qquad W^{s\pm}  \ , \qquad W^{s1} 
\ee
of spin $s$, $s+\frac{1}{2}$, and $s+1$, respectively. 
It would be interesting to characterise this twisted representation again more abstractly, i.e., as some sort of 
special level-half representation, and this is currently under investigation \cite{DGGK}.

\subsubsection{The Case with ${\cal N}=4$ Superconformal Symmetry}

In the ${\cal N}=4$ case, the analogue of (\ref{twisN2}) is 
\be\label{twisN4}
\chi_{\rm tw}^{{\cal N}=4}  (q,y) = \prod_{n=1}^{\infty} \frac{\bigl( 1 + y q^{n-\frac{1}{2} -\nu} \bigr)^2  \, \bigl( 1 + y^{-1} q^{n-\frac{1}{2} +\nu} \bigr)^2}
{\bigl( 1 - y q^{n-\nu} \bigr)^2 \, \bigl( 1 - y^{-1} q^{n-1 -\nu} \bigr)^2} \ . 
\ee
There are then different natural twisted representations. For example, taking the $y^0$ or $y^2$ coefficient of 
(\ref{twisN4}), we get the wedge characters
\be
\chi_0^{{\cal N}=4}(q)  = 
\chi_2^{{\cal N}=4}(q) = q^h \Bigl( 1 + 4 q^{1/2} + 11 q^1 + 28 q^{3/2} + 66 q^{2} + 144 q^{5/2} + 296 q^3 + \cdots \Bigr)
\ee
which agree precisely with the wedge character of the coset representations
\be
\Bigl( [m,0,\ldots,0];[m,0,\ldots,m],2m \Bigr) \ \hbox{or} \
\Bigl( [m,0,\ldots,0];[m+2,0,\ldots,m],2m+2(N+2) \Bigr) \ , 
\ee
see \cite{Gaberdiel:2014cha}, while the $y^1$ coefficient equals
\be\label{chi1}
\chi_1^{{\cal N}=4}(q) = q^h \Bigl( 2 + 4 q^{1/2} + 12 q^1 + 32 q^{3/2} + 74 q^{2} + 156 q^{5/2} + 316 q^3 + \cdots \Bigr) \ ,
\ee
and agrees precisely with the wedge character of the coset representation
\be
\Bigl( [m,0,\ldots,0];[m+1,0,\ldots,m],2m +(N+2) \Bigr)  \ ,
\ee
see again \cite{Gaberdiel:2014cha}. It would be interesting to repeat the above considerations for this case, and
in particular, characterise these representations more abstractly, see again \cite{DGGK} for further details.

\section{The Twisted Sector from the HSS viewpoint}\label{sec:twisHSS}

Let us now turn to the description of the twisted sector from the HSS viewpoint. As before, we shall
concentrate on the bosonic case for simplicity. Since the HSS algebra contains charged,
as well as uncharged operators, see eq.~(\ref{bosHSS}),
it does not make sense to decompose the individual $\nu$-twisted contributions into 
their charge components as in (\ref{charnu}); instead, we should work directly with the characters (\ref{chinu}).
It is furthermore clear from the structure of the HSS algebra that, as a vector space, the full twisted sector 
(say of the $m$-cycle twist) 
is a single irreducible representation of the HSS algebra whose character equals the product of the $\nu$-twisted
characters (\ref{chinu}) with $\nu=0,\frac{1}{m},\ldots,\frac{m-1}{m}$. 

The corresponding representation of the HSS has therefore an exponential (i.e., Cardy-like) growth, just as for the 
case considered in the previous section, see Sec.~\ref{sec:4.2}.
However, unlike the situation considered there, this is now the {\em same} as for the (tensor powers of the) minimal 
representation of the HSS, since the HSS minimal representation also has a similar growth behaviour, see eq.~(\ref{HSSmin}). 
From the viewpoint of the HSS, the representations that appear in the untwisted and twisted sector are thus on a similar footing --- 
there is no marked distinction in the size of the representations. 
This is in line with the fact that
in string theory there are perturbative (supergravity) states in both the untwisted and twisted sectors. 
On the other hand, this is in contrast to the situation for the higher spin algebra where, as discussed in previous sections, 
the twisted sector representations are distinctly larger than the corresponding minimal representation and hence 
leads to a distinction between `perturbative' vs.\ `non-perturbative states'. 

\smallskip

While this gives a good description of the twisted representation as a vector space, the structure as a representation
of the HSS is somewhat more complicated. Indeed, as was already alluded to before, the analysis cannot be split up 
any longer into the different $\nu$-twisted components. For example, if we consider the action of the 
neutral spin $4$ generator (\ref{Vgen}), then the four modes may in fact come from different $\nu$-twisted sectors ---
the only condition that has to be satisfied is that the total sum of mode numbers is again an integer, but there
are many ways to achieve this. A more fancy way to say this is that the Hopf structure (in particular
the comultiplication formula) for the HSS algebra is more complicated. This may be related to the fact that
the HSS plays, in some sense, the role of the Yangian for this system. 

Because of this difficulty, we are not able, at present, to describe this representation very explicitly for the
case of a general $m$-cycle twisted sector; however, we can be fairly concrete for the $2$-cycle twisted
case, for which we just have untwisted $\nu=0$ and $\nu=\frac{1}{2}$-twisted modes.

\subsection{The $2$-cycle twisted case}

To get a sense of the structure of the resulting representation, we have worked out the modes of $V$ (see eq.~({\ref{Vgen})) in the 
$2$-cycle twisted sector corresponding to the $2$-cycle $(12)$.  It is useful to split the modes into those two modes that are affected
by the twist, namely
\be\label{al}
\alpha_m = \frac{1}{\sqrt{2}} \, \bigl( \alpha^1_m + \alpha^2_m\bigr) \ , \qquad
\bar\alpha_m = \frac{1}{\sqrt{2}} \, \bigl( \bar\alpha^1_m + \bar\alpha^2_m\bigr) 
\ee
for $m\in\mathbb{Z}$, and 
\be\label{alT}
\alpha^{({\rm T})}_r = \frac{1}{\sqrt{2}} \, \bigl( \alpha^1_r - \alpha^2_r\bigr) \ , \qquad
\bar\alpha^{({\rm T})}_r = \frac{1}{\sqrt{2}} \, \bigl( \bar\alpha^1_r - \bar\alpha^2_r\bigr) 
\ee
for $r\in\mathbb{Z}+\frac{1}{2}$,\footnote{These modes then satisfy the commutation relations
$[\alpha_m,\bar\alpha_n]= m \delta_{m,-n}$, $[\alpha_m,\bar\alpha^{({\rm T})}_r]=0$, 
$[\alpha^{({\rm T})}_r,\bar\alpha^{({\rm T})}_s]=r \delta_{r,-s}$, etc.} 
 and the remaining $(N-2)$ complex modes associated to the directions $i=3,\ldots, N$.
Since the centraliser of the (12)-cycle $\mathbb{Z}_2\times S_{N-2}$ does not mix the two sets of modes,
we can separately define single-trace generators associated to (\ref{al}) and (\ref{alT}) on the one hand, and to the remaining 
modes on the other. Let us denote the Virasoro and $V^{(4)}$ (the superscript now making the spin of this current explicit) modes 
associated to (\ref{al}) and (\ref{alT}) by 
$L^\sigma_m$ and $V^{(4)\, \sigma}_m$, respectively, while the modes associated to the remaining directions
will be denoted by $L^\perp_m$ and $V^{(4)\, \perp}_m$, respectively. Then these modes satisfy the respective commutation
relations
\be\label{commsig}
[L^{\sigma} _m,V^{(4)\, \sigma}_{n} ] = (3m-n) V^{(4)\, \sigma}_{m+n}  + \frac{4}{3} m\,  (m^2-1)\, L^{\sigma}_{m+n} 
\ee
and
\be
[L^{\perp} _m,V^{(4)\, \perp}_{n} ] = (3m-n) V^{(4)\, \perp}_{m+n}  + \frac{2N-4}{3} m\,  (m^2-1)\, L^{\perp}_{m+n}  \ .
\ee
Note that the `full' Virasoro and $V^{(4)}$ generators are then simply 
\be
L_m = L^\sigma_m + L^\perp_m \ , \qquad 
V^{(4)}_m =  V^{(4)\, \sigma}_{m} + V^{(4)\,\perp}_m + 4 \, \sum_{l} L^\sigma_l\,  L^\perp_{m-l} \ ;
\ee
indeed, these modes then satisfy the commutation relation eq.~(\ref{LVcom}), where we have used that the 
$L^\sigma$ and $L^\perp$ generators define Virasoro algebras with central charge $c=4$ and $c=(2N-4)$, respectively. 
The formula for the $V^{(4)\, \perp}$ generators is as in the untwisted sector, see eq.~(\ref{Vmod}), where the 
invisible $i$-index now runs over $i=3,\ldots, N$. In the following we shall therefore only consider the 
$V^{(4)\, \sigma}$ generators. 

Since $V$ is quartic, there are a number of different contributions to $V^{(4)\, \sigma}_m$, which we denote by 
$V^{(4)[a,\bar{a}]}_m$, where $a$ and $\bar{a}$ count the number of unbarred and barred twisted modes. 
More specifically, we have 
\be\label{5.1}
V^{(4)\,\sigma}_m =  V^{(4)[0,0]}_m + V^{(4)[2,2]}_m + 4  \, V^{(4)[1,1]}_m 
+  V^{(4)[2,0]}_m +  V^{(4)[0,2]}_m \ . 
\ee
The  modes $V^{(4)[0,0]}_m$ take the same form as in (\ref{Vmod}), and the same is true for $V^{(4)[2,0]}_m$ and
$V^{(4)[0,2]}_m$, with the only modification that the sums over $p$, $r$, and $s$ run only over the appropriate values.
The modes $V^{(4)[1,1]}_m$ and $V^{(4)[2,2]}_m$, on the other hand, pick up normal-ordering terms, and we find 
explicitly that 
\be
V^{(4)[1,1]}_m = \sum_{p\in\mathbb{Z}} \sum_{r,s\in\mathbb{Z}+\frac{1}{2}}\, :\alpha_{p} \, \alpha^{({\rm T})}_{r} \, 
\bar\alpha^{({\rm T})}_{s} \, \bar\alpha_{m-p-r-s} :  
+ \frac{1}{8}\, \sum_{n\in\mathbb{Z}} :\alpha_n \bar\alpha_{m-n}:  \ , 
\ee
and
\be\label{5.3}
V^{(4)[2,2]}_m = \sum_{t,r,s\in\mathbb{Z}+\frac{1}{2}}\, :\alpha^{({\rm T})}_{t} \, \alpha^{({\rm T})}_{r} \, \bar\alpha^{({\rm T})}_{s} \, 
\bar\alpha^{({\rm T})}_{m-t-r-s} : 
+  \frac{1}{2}\, \sum_{r\in\mathbb{Z}+\frac{1}{2}} :\alpha^{({\rm T})}_r \bar\alpha^{({\rm T})}_{m-r}:  + \frac{1}{32} \, \delta_{m,0} \ .
\ee
Here we have imposed that $V^{(4)\, \sigma}_m$ has to satisfy (\ref{commsig}) with respect to the Virasoro generators 
\be\label{5.4}
L^\sigma_m = L^{(2)[0,0]}_m + L^{(2)[1,1]}_m =  \sum_{n\in\mathbb{Z}} \, : \alpha_{n} \bar\alpha_{m-n}: 
+ \sum_{r\in\mathbb{Z}+\frac{1}{2}}\, 
: \alpha^{({\rm T})}_r \bar\alpha^{({\rm T})}_{m-r} : + \frac{1}{8}\, \delta_{m,0} \ , 
\ee
see eq.~(\ref{Ltwist}) with $\nu=\frac{1}{2}$. 
As a consistency check we note 
that the eigenvalue of $V^{(4)}_0$ on the twisted sector ground state equals --- only the last term from eq.~(\ref{5.3}) 
contributes
\be\label{veigen}
v =  \frac{1}{32} \ . 
\ee
On the other hand, this is in agreement with the null-vector relation (\ref{WsN1}), since 
for $V^{(4)}_{-1} \, \phi$, the only term that contributes comes from (\ref{5.3})
\be
V^{(4)}_{-1} \, \phi = \frac{1}{2} \, \alpha^{({\rm T})}_{-1/2}\, \bar\alpha^{({\rm T})}_{-1/2} \, \phi 
= \frac{2v}{h}\, L^{\sigma}_{-1}\, \phi \ ,
\ee
where we have used that $h=\frac{1}{8}$, as follows from eq.~(\ref{5.4}). 
Since the structure of (\ref{5.1}) is quite complicated (and mixes different twisted sectors), it is not straightforward
to analyse the action of the $V^{(4)}_m$ modes sector by sector.

\subsection{Constraints on the BPS Spectrum}

Another natural question to ask is to which extent the Higher Spin Square (or indeed the higher spin ${\cal W}_\infty$
algebra) 
constrains the BPS spectrum of the symmetric orbifold theory. As is explained in \cite{David:2002wn}, the single-particle 
(i.e., single-cycle) BPS states of the symmetric orbifold arise from the $k$'th twisted sector, where in addition to the BPS 
progeny at $h=j=(k-1)/2$, there are $4$ BPS progenies at $h=j=k/2$ and one BPS progeny at $h=j=(k+1)/2$. 
Here `BPS progeny' means that these BPS states arise in the same twisted sector, i.e., are descendants under the action 
of the twisted modes of the twisted sector ground state (which is not BPS, except for the special case of $k=2$). The 
above issue then amounts to the question of whether 
these progenies are in the same (or in different) representations of the Higher Spin Square. 

Let us denote, as in \cite{David:2002wn}, the BPS descendant at $h=j=(k-1)/2$ as $\Sigma^{(k-1)/2}$. Then the 
$4$ BPS progenies at $h=j=k/2$ are schematically of the form
\be\label{5.12}
\sum_{A} \psi^{i}_A \tilde{\psi}^j_{A} \, \Sigma^{(k-1)/2} \ , 
\ee
where the sum over $A$ guarantees that the resulting state is permutation invariant, and $\psi^i$ and $\tilde{\psi}^j$
are left- and right-moving fermions. (Here $i$ and $j$ only run over two values each; the other two values do not give
rise to BPS states.)

It is now not difficult to see that these states are not in the same representation of the Higher Spin Square: while
$\Sigma^{(k-1)/2}$ is a singlet under the permutation action (both on the left and the right), the states in eq.~(\ref{5.12}) 
transform in the standard representation of the permutation group (separately left- and right), and only the left-right-combined
state is permutation invariant. Thus this state cannot be a descendant under the action of the Higher Spin Square (since
the latter only consists of generators that are permutation invariant). The situation is essentially the same for the 
progeny at $h=j=(k+1)/2$, which is schematically of the form 
\be
\sum_{A} \psi^{1}_A \psi^{2}_A  \tilde{\psi}^1_{A}  \tilde{\psi}^2_{A} \, \Sigma^{(k-1)/2} \ , 
\ee
which again transforms in a different representation of the permutation group --- the antisymmetric product of the
standard representation --- and hence cannot be a descendant of the Higher Spin Square. Thus the Higher Spin Square
representation theory, unfortunately, does not restrict the structure of the BPS states.

\section{Conclusions}

In this paper we have analysed the CFT dual of string theory on AdS$_3\times {\rm S}^3 \times \mathbb{T}^4$ --- the symmetric
orbifold of $\mathbb{T}^4$ --- through the lens of a higher spin theory. From this viewpoint, the relevant
higher spin symmetry algebra is the Higher Spin Square (HSS), and the full spectrum can be organised in terms of 
representations of this algebra. In particular, we have seen that the untwisted sector of the symmetric orbifold CFT
consists precisely of the multiparticle states of the minimal HSS representation (as well as the higher spin descendants); 
thus it has exactly the same structure as the perturbative part of the CFTs that appear in the usual minimal model holography
(and that describe the higher spin and massive scalar degrees of freedom of the dual higher spin theory). This 
suggests that the whole untwisted sector of the symmetric orbifold CFT is dual to a Vasiliev type higher spin theory
on AdS$_3$, with the only difference that the symmetry algebra (the HSS) is now much larger than a usual higher spin algebra.

{}From the viewpoint of the HSS, the twisted sector seems to have a similar structure in that it is built from 
finite tensor products of representations that are only slightly larger than the minimal representation. 
This is in line with the 
fact that the twisted sector of the symmetric orbifold CFT also describes perturbative string states, just as the untwisted sector. 
Thus from the viewpoint of the formulation of the higher spin theory (based on HSS) with matter couplings, these states 
should not be qualitatively different from the minimally coupled ones. It would be
very important to understand how exactly one adds them to the Vasiliev-like
HSS-based higher spin theory without breaking the higher spin symmetry, etc.\footnote{In this context it should be noted
that the conformal dimension of the twisted sector ground state depends on the length of the cycle, and hence
that the mass of these matter fields is apparently not fixed by the underlying symmetry algebra --- this is in 
contrast to what is believed to be the case for conventional higher spin theories based on ${\rm hs}[\lambda]$,  say, where the mass 
of the scalar field is believed to be uniquely fixed by $\lambda$, see, e.g., \cite{Prokushkin:1998bq}.}
The underlying gauge invariance must strongly 
constrain their couplings and it would be interesting to compare these to the predictions from the orbifold CFT. 

Our analysis has mainly been bosonic, but we have sketched the structure of the supersymmetric generalisation;
it would also be interesting to work this out in more detail, and we hope to return to this point in the near future \cite{DGGK}. 
Some of the features of the HSS (even in the bosonic case) are reminiscent of Yangians, as we mentioned in the context 
of the twisted sector representations. Also the existence of horizontal and vertical HS algebras generating the HSS is 
somewhat like the superconformal and the dual superconformal algebras generating the entire Yangian that underlies 
${\cal N}=4$ Super Yang-Mills. In this context, it is encouraging to note that the ${\cal W}_{\infty}[\lambda]$ algebras themselves 
can be mapped into the Yangian of $\hat{\mathfrak{gl}}(1)$ \cite{Prochazka}.

\section*{Acknowledgements}

We thank  Shouvik Datta, Justin David, Terry Gannon, Maximilian Kelm, Shiraz Minwalla, 
Misha Vasiliev, and Gerard Watts for useful discussions. One of us (R.G.) would like to thank the organisers of the workshop 
``All About ${\rm AdS}_3$" at ETH Zurich for hospitality while this work was completed. We would like to thank the participants 
of the workshop for stimulating conversations.

\appendix

\section{Some stringy commutators}\label{app:sa}

Recall from Sec.~\ref{sec:SA} that the lowest neutral generator of the Higher Spin Square that is not an element of
the original higher spin algebra is the field 
\be
V =  \sum_i \partial \phi^i \, \partial\phi^i \,\, \partial\bar\phi^i\, \partial\bar\phi^i\ , 
\ee
whose mode expansion was given in eq.~(\ref{Vmod}). One convenient way to make the normal ordering 
of $V_m$ explicit is to write (\ref{Vmod}) as 
\begin{eqnarray}
V_m & = & \sum_{p,r \leq 0}\, \sum_{s\in\mathbb{Z}} \alpha_{p} \, \alpha_{r} \, \bar\alpha_{s} \, \bar\alpha_{m-p-r-s} 
+ \ 2 \, \sum_{p\leq 0, r>0}\, \sum_{s\in\mathbb{Z}} \alpha_{p} \, \bar\alpha_{s} \, \bar\alpha_{m-p-r-s} \,  \alpha_{r}  \nonumber \\
&  & \quad + \sum_{p,r > 0}\, \sum_{s\in\mathbb{Z}} \bar\alpha_{s} \, \bar\alpha_{m-p-r-s} \,  \alpha_{p} \, \alpha_{r} \ .
\end{eqnarray}
Then, being careful about the normal ordering terms, one finds for the commutator
\be\label{LVcom}
{} [ L_n,V_m] = (3n- m ) V_{n+m} + \frac{2N}{3} n (n^2-1) \, L_{m+n} \ ,
\ee
where $N$ is the number of copies. In particular, the field $V$ is therefore not primary, but only quasi-primary. 
\smallskip

We have also worked out some of the low-lying commutators of this field with the ${\cal W}_\infty[1]$
generators, using the general formula for the operator product expansion of chiral fields 
\be
W^{(s)}(z) \, V(u) \sim \sum_{m=-s+1}^{4} \frac{(W^{(s)}_m V)(u)}{(z-u)^{m+s}} \ , 
\ee
where $(W^{(s)}_m V)(u)$ is the field associated to the state $W^{(s)}_m\, V$, and the latter can be worked out
using the mode expansions for the low-lying spin fields, see eqs.~(\ref{Wmode}) and (\ref{Umode}), as well as 
\be
V \equiv \alpha_{-1} \alpha_{-1} \bar\alpha_{-1} \bar\alpha_{-1} \, |0\rangle \ . 
\ee
Proceeding in this manner we find, for example, that (writing $W\equiv W^{(3)}$)
\be
W(z) \, V(u) \sim \frac{12 S^{(5)}(u)}{(z-u)^2} + \frac{\frac{24}{5} S^{(5)'}(u) + \frac{8}{5} T^{(6),0}(u)}{(z-u)} \ , 
\ee
where $S^{(5)}$ and $T^{(6),0}$ are the fields associated to the states
\begin{eqnarray}
S^{(5)} & = & \bigl( \alpha_{-2} \alpha_{-1} \bar\alpha_{-1}^2 - \bar\alpha_{-2} \bar\alpha_{-1} \alpha_{-1}^2 \bigr) \, |0\rangle \\
T^{(6),0} & = & \bigl( 3 \alpha_{-2}^2 \bar\alpha_{-1}^2 - 3  \bar\alpha_{-2}^2 \alpha_{-1}^2 
- 4 \alpha_{-3} \alpha_{-1}  \bar\alpha_{-1}^2 + 4 \bar\alpha_{-3} \bar\alpha_{-1}  \alpha_{-1}^2 \bigr)\, |0\rangle \ . 
\end{eqnarray}
In terms of modes, this then translates into the commutation relation
\be\label{WVcom}
{} [ W_m,V_n] = \frac{12}{5} (3m-2n)\, S^{(5)}_{m+n} + \frac{8}{5} T^{(6),0}_{m+n} \ . 
\ee
The commutator of the spin $3$ field $W$ with $S^{(5)}$ can be worked out using the same techniques, and one finds
\begin{eqnarray}
{} [W_m, S^{(5)}_n ] & = &  - \frac{1}{15} \bigl(-28 m^3 + 21 m^2 n + m (88-9n^2) + 2n (n^2-16) \bigr) V_{m+n} \nonumber \\
& & \quad + \frac{4}{3}\, (2m-n) \, T^{(6),1}_{m+n} + U^{(7)}_{m+n} \ , \label{WScom}
\end{eqnarray}
where the corresponding states are 
\begin{eqnarray}
T^{(6),1} & = & \bigl( \alpha_{-2} \bar\alpha_{-1}^2  + \bar\alpha_{-2} \alpha_{-1}^2 + 4 \alpha_{-3} \alpha_{-1} \bar\alpha_{-1}^2 
+ 4 \bar\alpha_{-3} \bar\alpha_{-1} \alpha_{-1}^2 - 8 \alpha_{-2} \bar\alpha_{-2} \alpha_{-1} \bar\alpha_{-1} \bigr) |0\rangle \quad \nonumber \\
U^{(7)} & = & \bigl( 3 \alpha_{-2}^2 \bar\alpha_{-2} \bar\alpha_{-1} + 3 \bar\alpha_{-2}^2 \alpha_{-2} \alpha_{-1}
- 4 \alpha_{-3} \bar\alpha_{-2} \alpha_{-1} \bar\alpha_{-1} - 4 \bar\alpha_{-3} \alpha_{-2} \bar\alpha_{-1} \alpha_{-1}  \nonumber \\
& & \quad 
- 2 \alpha_{-3} \alpha_{-2} \bar\alpha_{-1}^2  - 2 \bar\alpha_{-3} \bar\alpha_{-2} \alpha_{-1}^2  
+ 3 \alpha_{-4} \alpha_{-1} \bar\alpha_{-1}^2 + 3 \bar\alpha_{-4} \bar\alpha_{-1} \alpha_{-1}^2 \bigr) |0\rangle \ ,
\end{eqnarray}
while the commutator with $T^{(6),0}$ is 
\begin{eqnarray}
{} [W_m, T^{(6),0}_n ] & = &  
\frac{1}{165} \Bigl( - 840 m^4 + 336 m^3 n - 112 m^2 n^2 + 28 m n^3 - 4 n^4  \nonumber \\
& & \qquad \quad + 3640 m^2 - 1036 mn + 164 n^2 - 1600\Bigr)\, V_{m+n} \label{WTcom} \\ 
& &  + \frac{1}{13} \bigl( - 10 m^2 + 6mn - \frac{4}{3} n^2 + \frac{40}{3} \bigr) \bigl[ T^{(6),1}_{m+n} - 10 \, T^{(6),2}_{m+n} \bigr]
+ \frac{80}{1001} R^{(8)}_{m+n} \ , \quad  \nonumber
\end{eqnarray}
where 
\be
T^{(6),2} = \bigl(3\alpha_{-2}^2 \alpha_{-1}^2   + 3\bar\alpha_{-2}^2 \alpha_{-1}^2  
- 4 \alpha_{-3} \alpha_{-1} \bar\alpha_{-1}^2 - 4 \bar\alpha_{-3} \bar\alpha_{-1} \alpha_{-1}^2 \bigr)\, |0\rangle \ , 
\ee
and the expression for $R^{(8)}$ is too cumbersome to write down. Finally, the commutator of the spin $4$ field $U\equiv W^{(4)}$ with
$V$ can then be expressed in terms of these modes, and one finds 
\begin{eqnarray} \label{UVcom}
{}[U_m,V_n] & =  & \frac{32}{15} (m-n) \bigl( m^2 - mn + n^2 -7)\, V_{m+n}  \nonumber \\
& & + \frac{16}{3} (m-n)\, 
\bigl( T^{(6),1}_{m+n} - T^{(6),2}_{m+n} \bigr)+ \frac{64}{5}\, U^{(7)}_{m+n} \ . 
\end{eqnarray}

\section{The {hs[1]} Representation with one state at level one}\label{A.1}

In this appendix we analyse the structure of a generic level one representation of ${\rm hs}[1]$, i.e., of a 
representation with a single state at level one.  We will be completely general and not make any direct reference to any
specific realisation of such a representation (such as the $\nu$-twisted representation). 

Let us begin by analysing the level one representation at the first excited level. Since there is only one state at this level, 
it follows that 
\be\label{WsN1chi}
W^{(s)}_{-1} \chi = \frac{ s w^{(s)}}{ 2h} \,  L_{-1}\, \chi \ ,
\ee
where $\chi$ is the ground state of the level one representation. (Again, the proportionality constant is required by 
consistency with the algebra relations, i.e., follows from the application of $L_1$ to this identity.)

We now want to show that all the null-vectors (\ref{WsN1chi}) with $s\geq 4$ are
a consequence of the null-vector with $s=3$, and that this in turn fixes also all the eigenvalues $w^{(s)}$ 
with $s\geq 4$. The idea for this is rather simple. Suppose we have the null-vector
\be\label{level1gen}
W_{-1} \, \chi \equiv W^{(3)}_{-1} \, \chi = \alpha\, L_{-1}\, \chi \ .
\ee
[If $\chi$ is the ground state of the twisted representation, then 
$\chi=\phi$ and $\alpha= 2 (2\nu-1)$, but now we are considering
a general highest weight representation with highest weight $\chi$ and null-vector (\ref{level1gen}).] 
Then we apply $W_0\equiv W^{(3)}_0$ to both sides of this equation. Since $\chi$ is an eigenvector of $W_0$
with eigenvalue $w^{(3)} = \frac{2}{3} \alpha h$, see eq.~(\ref{WsN1chi}), we get the identity
\be
{}[W_0, W_{-1}] \, \chi = \alpha\, [W_0,L_{-1}]\, \chi = 2 \alpha\, W_{-1} \, \chi = 2 \alpha^2 \, L_{-1} \, \chi \ .
\ee
On the other hand, the left-hand-side can be evaluated using the commutation relations of ${\rm hs}[1]$, 
\be
{}[W_m,W_n] = n_0 (m-n) (2m^2 - mn +2n^2 - 8) L_{m+n} + 2 (m-n) U_{m+n} \ ,  \quad n_0 = \frac{12}{15}  \ ,
\ee
where $n_0=\frac{12}{15}$ is just a normalisation convention --- it fixes the normalisation of the spin $3$ $W$-field --- and the 
above value is  chosen so as to be in agreement with the normalisation convention of \cite{Gaberdiel:2013jpa}. 
This then leads to the null-vector relation 
\be\label{Uchi}
U_{-1}\, \chi = \bigl(\alpha^2 + \tfrac{12}{5} \bigr) \, L_{-1} \chi \ , 
\ee
which for the case of the twisted representation, i.e., $\alpha= 2 (2\nu-1)$, reproduces precisely (\ref{NU1}). Continuing
in this manner, we can work out similarly the null-vectors for $s=5$ and $s=6$, and they turn out to be of the form (\ref{WsN1}) with 
\begin{eqnarray}
w^{(5)} & = & 
\frac{2h}{5}\, \alpha \bigl(\alpha^2 + \frac{68}{7} \bigr) \ , \label{w5} \\
w^{(6)} & = & \frac{h}{21} \, \bigl(7 \alpha^4+168 \alpha^2+240\bigr) \ .  \label{w6}
\end{eqnarray}

Next we want to show that these null-vectors also imply that there are only $3$ states at level $2$, without any
additional assumptions. At level $2$ we can either have two $(-1)$-modes, or a single $(-2)$-mode. Because of the
null-vectors at level $1$, (\ref{WsN1chi}), we can replace the $(-1)$-mode acting on $\chi$ by $L_{-1}$, and then
commute the other $(-1)$ mode so as to act directly on $\chi$, where in turn it can again be replaced by a $L_{-1}$-mode,
using again (\ref{WsN1chi}). Thus the most general state at level $2$ is of the form
\be\label{basis2chi}
W^{(s)}_{-2} \, \chi  \ \ (s\geq 3) \ , \qquad L_{-1}^2 \, \chi \ . 
\ee
We now want to show that this space is only $3$-dimensional, i.e., that there are null-relations relating 
$W^{(5)}_{-2}\,\chi$ to $W^{(3)}_{-2} \, \chi$  and $W^{(4)}_{-2} \, \chi$. [Note that $W^{(2)}_{-2}\, \chi \equiv L_{-2} \, \chi$
is not generated by a wedge mode.] Once we have a null-vector of this form, the higher null-vectors can then again
be generated by acting with $W^{(3)}_0$ as above.

To obtain the relevant null-vector we apply $W_{-1}$ to the null vector (\ref{level1gen}), and obtain
\be\label{eq1}
W_{-1} W_{-1} \, \chi = \alpha\,\Bigl( W_{-2}\, \chi + \alpha\, L_{-1}^2 \, \chi\Bigr) \ , 
\ee
where we have used (\ref{level1gen}) again on the right-hand-side. Similarly, we get from (\ref{Uchi})
\be\label{WU}
W_{-1} U_{-1} \, \chi = \Bigl( \alpha^2 + \frac{12}{5}\Bigr) \,  \Bigl( W_{-2}\, \chi + \alpha L_{-1}^2 \, \chi\Bigr) \ , 
\ee
while 
\begin{eqnarray}
U_{-1} W_{-1}\, \chi & = & \alpha\,\Bigl[ 2 U_{-2}\, \chi +   \Bigl( \alpha^2 + \frac{12}{5}\Bigr) \,  L_{-1}^2 \, \chi \Bigr] \nonumber \\
& = & - \tfrac{384}{35}\, W_{-2}\, \chi + X_{-2} \, \chi  + W_{-1}  U_{-1}\, \chi\ . \label{eq3}
\end{eqnarray}
In going to the second line we have used the (wedge) commutation relations
\be\label{WUcom}
{}[W_m,U_n] = n_3 \bigl(-5 m^3 + 5 m^2n + m(17-3n^2)+n^3-9n\bigr) W_{m+n}  + (3m-2n) X_{m+n}  \ , 
\ee
where $n_3 = - \tfrac{64}{35}$, see, e.g., \cite{Gaberdiel:2013jpa}.
Using (\ref{WU}), we can solve this expression for $X_{-2}\, \phi\equiv W^{(5)}_{-2}\, \phi$ in terms of the basis vectors (\ref{basis2chi}), and hence obtain
\begin{equation}\label{X2}
W^{(5)}_{-2}\, \chi = 2 \alpha \, U_{-2}\, \chi - \bigl( \alpha^2 - \tfrac{60}{7} \bigr) W_{-2}\, \chi \ .
\ee
Note that this reproduces, upon setting $\alpha=2(2\nu-1)$, precisely the null-vector of eq.~(\ref{level2N}).

A similar argument allows us to show that the vector space of non-trivial states at level $3$ is 
$6$-dimensional. At level $3$, the possible descendants 
are either a single $(-3)$-mode, or the product of a $(-2)$- and $(-1)$-mode, or the product of three $(-1)$-modes. 
Since the $(-1)$-modes on $\chi$ are all proportional to $L_{-1}$ (and since we can, using commutators, always move
the $(-1)$-mode to the right so as to act on $\chi$), it follows that all $(-1)$-modes are $L_{-1}$. By the same token, 
the $(-2)$-modes are either
$W_{-2}$ or $U_{-2}$, since we have (\ref{X2}), and hence all $W^{(s)}_{-2}\, \chi$ with $s\geq 5$ can be expressed in terms
of $W_{-2}\,\chi$ and $U_{-2}\,\chi$. To see which states with $(-3)$ modes exist, we would like to apply $W_{-1}\equiv W^{(3)}_{-1}$ 
to the relation (\ref{X2}), but because of the structure of the commutators
\be\label{Wcomm}
{}[W^{(3)}_{m}, W^{(s)}_{n}]  \propto \bigl( (s-1)m - 2 n \bigr)\, W^{(s+1)}_{m+n} + \hbox{lower spin terms} \ , 
\ee
the leading term (proportional to $W^{(6)}_{-3}\chi$) vanishes. Thus we apply instead $W_{-1}$ to the relation expressing
$W^{(6)}_{-2}\chi$ in terms of $U_{-2}\chi$ and $W_{-2}\chi$, 
\begin{eqnarray}
W_{-1} W^{(6)}_{-2} \chi & = & [W_{-1}, W^{(6)}_{-2}] \, \chi + W^{(6)}_{-2} W_{-1} \chi \nonumber \\
& = & \# \, W^{(7)}_{-3} \chi + \hbox{(lower spin terms)} + \# \, W^{(6)}_{-2} \, L_{-1}\, \chi \nonumber \\
& = & \# \, W^{(7)}_{-3} \chi + \hbox{(lower spin terms)} +  \# \, L_{-1} W^{(6)}_{-2} \, \chi  \\
& = & \# \, W^{(7)}_{-3} \chi + \hbox{(lower spin terms)} +  \# \, L_{-1} W_{-2} \, \chi +  \# \, L_{-1} U_{-2} \, \chi  \ , \nonumber
\end{eqnarray}
where, in the last line, we have used (\ref{X2}) again. Performing a similar manipulation on the right-hand-side
then leads to an identity relating $W^{(7)}_{-3} \chi $ in terms of lower spin $(-3)$ modes, i.e., there are only
three linearly independent $(-3)$ modes acting on $\chi$, namely $U_{-3}$, $X_{-3}$, and $Y_{-3}\equiv W^{(6)}_{-3}$. 
Together with the other three states, namely $L_{-1} W_{-2}\chi$, $L_{-1} U_{-2}\chi$ and $L_{-1}^3 \chi$, 
this therefore gives rise to $6$ linearly independent states. 

The analysis at the next level works similarly. Apart from the products of lower modes we need to constrain
the possible $(-4)$ modes, and since the leading term in the commutator $[W_{-1},W^{(7)}_{-3}]$
vanishes because of (\ref{Wcomm}), we need to apply $W_{-1}$ to the null-vector expressing $W^{(8)}_{-3}\chi$
in terms of lower modes. Thus we find four $(-4)$ modes,
namely those associated to spin $s=5,6,7,8$. This pattern continues at higher levels, and thus the character of the
level-one representation is the generic Verma module character where one has $n$ modes of type $(-n)$. So at
level $4$, the relevant counting gives 
\be
\hbox{level 4:}  \qquad \qquad 
\begin{array}{ll}
\hbox{mode shape $(-4)$} \qquad & \hbox{$4$ states}  \\
\hbox{mode shape $(-3)(-1)$} \qquad & \hbox{$3$ states} \\
\hbox{mode shape $(-2)^2$} \qquad & \hbox{$3$ states}  \\
\hbox{mode shape $(-2)(-1)^2$} \qquad & \hbox{$2$ states}  \\
\hbox{mode shape $(-1)^4$} \qquad & \hbox{$1$ state} 
\end{array}
\ee
for a total of $13$ states, while at level $5$ we obtain similarly $24$ states. 
In fact, the wedge character of the generic level-one representation equals precisely
the MacMahon function
\be\label{MM}
\chi_{\rm near-min}^{({\rm wedge}), {\rm bos}}(q) = q^h\, \prod_{n=1}^{\infty} \frac{1}{(1-q^n)^n} = 
q^h \Bigl( 1 + q + 3 q^2 + 6 q^3 + 13 q^4 + 24 q^5 + 48 q^6 + \cdots \Bigr) \ . 
\ee

\subsection{The Relation at level four}\label{sec:twisnm}

Comparing (\ref{MM}) to (\ref{chinu}) we conclude that a generic level-one representation has more
states than those that characterise the $\nu$-twisted sector --- the characters differ at order $q^4$.
The origin of this null-vector
can be understood as follows. The level-one representation is generated by $n$ modes with
mode number $(-n)$, but from the viewpoint of the twisted sector, each such mode is a bilinear of 
the bosons. Thus there can be a relation between the product of two such modes; indeed, the first relation
of this kind appears at conformal weight $h=4$ where we have the identity
\be
\bigl( \alpha_{-2+\nu} \bar\alpha_{-\nu} \bigr) \, \bigl( \alpha_{-1+\nu} \bar\alpha_{-1-\nu} \bigr) \phi = 
\bigl( \alpha_{-2+\nu} \bar\alpha_{-1-\nu} \bigr) \, \bigl( \alpha_{-1+\nu} \bar\alpha_{-\nu} \bigr) \phi \ , 
\ee
i.e., one state of mode shape $(-2)^2$ equals one state of mode shape $(-3)(-1)$. This explains why
the actual twisted character has only $12$ states at level $4$, while the generic level-one representation
has $13$. 

In turn this implies that the twisted sector representation is a special case of a level-one representation that
is characterised, in addition to having only a single state at level $1$ by the condition of having a non-trivial
null-vector at level $4$. One would expect that the existence of this null-vector is only possible if 
$h$ and $w^{(3)}$ are related as in (\ref{twisquan}), and thus this additional condition may characterise
the $\nu$-twisted representation uniquely. (It may, however, also be the case that yet another condition will
be needed to single out the $\nu$-twisted representations.)

\subsection{$\nu$-twisted Representations from the Coset Viewpoint}\label{sec:nearmincos}

We can also understand the structure of the level-one representations from the coset viewpoint, in 
particular, that there is a $2$-parameter family of such representations that are 
characterised by the values of $h$ and $w^{(3)}$. Recall that, as explained in \cite{Niedermaier:1991cu} (see also
\cite{Bouwknegt:1992wg}, section 6.4.2), the completely degenerate coset  representations labelled 
by $(\Lambda_+;\Lambda_-)$ have $(N-1)$ null vectors that appear at levels
\be
(\Lambda_+ + \rho,\alpha_i) \, (\Lambda_- + \rho,\alpha_i) \ , \qquad i=1,\ldots, N-1 \ , 
\ee
where $\alpha_i$ are the fundamental roots, and $\rho$ is the Weyl vector. A level-one representation,
i.e., one for which $N-2$ of these $N-1$ null-vectors appear at level $1$, is characterised by the condition that both
$\Lambda_+$ and $\Lambda_-$ have the same non-vanishing Dynkin label. In particular, since the two non-vanishing
Dynkin labels need not be equal, there is a $2$-parameter family of such representations, matching the 
expectations from above. We have also calculated the characters of these representations,\footnote{We thank
Maximilian Kelm for providing the relevant Mathematica notebook for us.} and for generic
values of the Dynkin labels they agree (to the extent to which we have checked this, i.e., up to order
$q^5$) with (\ref{MM}).

\bibliographystyle{JHEP}

\begin{thebibliography}{99}

\bibitem{Gross:1988ue} 
D.J.~Gross,
``High-energy symmetries of string theory,''
Phys.\ Rev.\ Lett.\  {\bf 60} (1988) 1229.
  
\bibitem{Witten:1988zd} 
E.~Witten,
``Space-time and topological orbifolds,''
Phys.\ Rev.\ Lett.\  {\bf 61} (1988) 670.
  
\bibitem{Moore:1993qe} 
G.W.~Moore,
``Symmetries and symmetry breaking in string theory,''
in proceedings of the SUSY '93 conference, `Supersymmetry and unification of fundamental interactions,'
(1993) 540 {\tt [arXiv:hep-th/9308052]}.
 
 \bibitem{Sagnotti:2011qp} 
A.~Sagnotti,
``Notes on strings and higher spins,''
J.\ Phys.\ A {\bf 46} (2013) 214006 
{\tt [arXiv:1112.4285 [hep-th]]}.

\bibitem{Gaberdiel:2014cha} 
M.R.~Gaberdiel and R.~Gopakumar,
``Higher Spins \& Strings,''
JHEP {\bf 1411} (2014) 044 
{\tt [arXiv:1406.6103 [hep-th]]}.

\bibitem{Gaberdiel:2015uca} 
M.R.~Gaberdiel, C.~Peng, and I.G.~Zadeh,
``Higgsing the stringy higher spin symmetry,''
JHEP {\bf 1510} (2015) 101 
{\tt [arXiv:1506.02045 [hep-th]]}.

\bibitem{Vasiliev:2003ev}
M.A.~Vasiliev,
``Nonlinear equations for symmetric massless higher spin fields in (A)dS(d),"
Phys.\ Lett.\  B {\bf 567} (2003) 139
{\tt [arXiv:hep-th/0304049]}.

\bibitem{Sundborg:2000wp}
B.~Sundborg,
``Stringy gravity, interacting tensionless strings and massless higher spins,"
Nucl.\ Phys.\ Proc.\ Suppl.\  {\bf 102} (2001) 113
{\tt [arXiv:hep-th/0103247]}.
  
\bibitem{Witten}
E.~Witten, talk at the John Schwarz 60-th birthday symposium (Nov. 2001), \newline
{\tt http://theory.caltech.edu/jhs60/witten/1.html}.

\bibitem{Mikhailov:2002bp}
A.~Mikhailov,
``Notes on higher spin symmetries,''
{\tt arXiv:hep-th/0201019}.

\bibitem{Gaberdiel:2013vva}
M.R.~Gaberdiel and R.~Gopakumar,
``Large $\mathcal{N}=4$ holography,''
 JHEP {\bf 1309} (2013) 036
{\tt  [arXiv:1305.4181 [hep-th]]}.

\bibitem{Gaberdiel:2010pz}
M.R.~Gaberdiel and R.~Gopakumar,
 ``An AdS$_3$ dual for minimal model CFTs,''
Phys.\ Rev.\ D {\bf 83} (2011) 066007
 {\tt [arXiv:1011.2986 [hep-th]]}.

\bibitem{Gaberdiel:2012uj}
M.R.~Gaberdiel and R.~Gopakumar,
``Minimal model holography,''
J.\ Phys.\ A: Math.\ Theor.\ {\bf 46} (2013) 214002
{\tt [arXiv:1207.6697 [hep-th]]}.

\bibitem{Gaberdiel:2015mra} 
M.R.~Gaberdiel and R.~Gopakumar,
``Stringy Symmetries and the Higher Spin Square,''
J.\ Phys.\ A {\bf 48} (2015) 185402 
{\tt [arXiv:1501.07236 [hep-th]]}.

\bibitem{Gaberdiel:2011zw}
M.R.~Gaberdiel, R.~Gopakumar, T.~Hartman, and S.~Raju, 
``Partition functions of holographic minimal models,''
JHEP {\bf 1108} (2011) 077
{\tt [arXiv:1106.1897 [hep-th]]}.

\bibitem{Castro:2011iw}
A.~Castro, R.~Gopakumar, M.~Gutperle, and J.~Raeymaekers,
``Conical defects in higher spin theories,''
JHEP {\bf 1202} (2012) 096
{\tt [arXiv:1111.3381 [hep-th]]}.

\bibitem{Gaberdiel:2012ku}
M.R.~Gaberdiel and R.~Gopakumar,
``Triality in minimal model holography,''
JHEP {\bf 1207} (2012) 127 
{\tt [arXiv:1205.2472 [hep-th]]}.

\bibitem{Perlmutter:2012ds}
E.~Perlmutter, T.~Prochazka, and J.~Raeymaekers,
``The semiclassical limit of W$_N$ CFTs and Vasiliev theory,''
JHEP {\bf 1305} (2013) 007
{\tt [arXiv:1210.8452 [hep-th]]}.

\bibitem{Bakas:1990ry} 
I.~Bakas and E.~Kiritsis,
``Boosnic realisation of a universal W algebra and $\mathbb{Z}_{\infty}$ parafermions,"
Nucl.\ Phys.\ B {\bf 343} (1990) 185
[Erratum ibid.\ B {\bf 350}, 512 (1991)].

\bibitem{Gaberdiel:2013jpa}
M.R.~Gaberdiel, K.~Jin, and W.~Li,
``Perturbations of ${\cal W}_\infty$ CFTs,''
JHEP {\bf 1310} (2013) 162
{\tt [arXiv:1307.4087 [hep-th]]}.

\bibitem{BBF}
V.~Bekkert, G.~Benkart, and V.~Futorny,
``Weyl algebra modules,"
{\tt arXiv:math/0202222 [math.RA]}.

\bibitem{Dijkgraaf:1996xw}
R.~Dijkgraaf, G.W.~Moore, E.P.~Verlinde, and H.L.~Verlinde,
``Elliptic genera of symmetric products and second quantized strings,''
Commun.\ Math.\ Phys.\  {\bf 185} (1997) 197
{\tt  [arXiv:hep-th/9608096]}.

\bibitem{Dutta:2007ws}
S.~Dutta and R.~Gopakumar,
``Free fermions and thermal AdS/CFT,''
JHEP {\bf 0803} (2008) 011
{\tt [arXiv:0711.0133 [hep-th]]}.

\bibitem{Candu:2012tr}
C.~Candu and M.R.~Gaberdiel,
``Duality in N=2 Minimal Model Holography,''
JHEP {\bf 1302} (2013) 070
{\tt [arXiv:1207.6646 [hep-th]]}.

\bibitem{Gaberdiel:2014vca} 
M.R.~Gaberdiel and M.~Kelm,
``The continuous orbifold of $ \mathcal{N} = 2$ minimal model holography,''
JHEP {\bf 1408} (2014) 084 
{\tt [arXiv:1406.2345 [hep-th]]}.

\bibitem{Fredenhagen:2012bw}
S.~Fredenhagen and C.~Restuccia,
``The geometry of the limit of N=2 minimal models,''
J.\ Phys.\ A {\bf 46} (2013) 045402
{\tt [arXiv:1208.6136 [hep-th]]}.

\bibitem{DGGK}
S.~Datta, M.R.~Gaberdiel, R.~Gopakumar, and M.~Kelm,
in preparation.

\bibitem{David:2002wn} 
 J.R.~David, G.~Mandal, and S.R.~Wadia,
``Microscopic formulation of black holes in string theory,''
Phys.\ Rept.\  {\bf 369} (2002) 549 
{\tt  [arXiv:hep-th/0203048]}.  

\bibitem{Prokushkin:1998bq}
S.F.~Prokushkin and M.A.~Vasiliev,
``Higher spin gauge interactions for massive matter fields in 3-D AdS space-time,''
Nucl.\ Phys.\ B {\bf 545} (1999) 385
{\tt [arXiv:hep-th/9806236]}.

\bibitem{Prochazka}
T.~Prochazka,
``${\cal W}$-symmetry, topological vertex and affine Yangian,"
preprint.

\bibitem{Niedermaier:1991cu} 
M.~Niedermaier,
``Irrational free field resolutions for W(sl(n)) and extended Sugawara construction,''
Commun.\ Math.\ Phys.\  {\bf 148} (1992) 249.

\bibitem{Bouwknegt:1992wg}
P.~Bouwknegt and K.~Schoutens,
``W symmetry in conformal field theory,''
Phys.\ Rept.\  {\bf 223} (1993) 183
{\tt [arXiv:hep-th/9210010]}.

\end{thebibliography}

\end{document}